\documentclass[fleqn,usenatbib]{mnras}

\usepackage{newtxtext,newtxmath}

\usepackage[T1]{fontenc}
\usepackage{ae,aecompl}

\usepackage{graphicx}	
\usepackage{amsmath}	

\usepackage{amssymb}	

\newcommand{\afe}{[$\alpha$/{\rm Fe}]}
\newcommand{\alf}{$\alpha$}
\newcommand{\ZH}{[Z/H]}
\newcommand{\re}{R$_e$}
\newcommand{\sgm}{$\sigma$}
\newcommand{\sgmRe}{$\sigma_{e}$}
\newcommand{\sgmReKms}{\sgmRe/km s$^{-1}$}
\newcommand{\Msun}{M$_\odot$}

\newcommand{\ppxf}{\texttt{pPXF}}
\newcommand{\mgb}{Mg$b$}
\newcommand{\mgI}{Mg$_1$}
\newcommand{\Rmaj}{R$_e^{maj}$}

\title[The \afe\ of MaNGA ETGs]{SDSS-IV MaNGA: The \afe\ of Early-Type Galaxies}

\author[Y. Liu]{
Yiqing Liu \thanks{E-mail: yiqing.liu@physics.ox.ac.uk, halfxianr@gmail.com}
\\
Sub-department of Astrophysics, Department of Physics, University of Oxford, Denys Wilkinson Building, Keble Road, Oxford OX1 3RH, UK\\
}

\date{Accepted XXX. Received YYY; in original form ZZZ}

\pubyear{2019}

\begin{document}
\label{firstpage}
\pagerange{\pageref{firstpage}--\pageref{lastpage}}
\maketitle

\begin{abstract}
The mean stellar alpha-to-iron abundance ratio (\afe) of a galaxy is an indicator of galactic star formation timescale. It is important for understanding the star formation history of early-type galaxies (ETGs) as their star formation processes have basically stopped. 
Using the model templates which are made by Vazdekis et al., we apply the \ppxf\ based spectral fitting method to estimate the \afe\ of 196 high signal-to-noise ratio ETGs from the MaNGA survey. The velocity dispersions within 1\re\ (\sgmRe) range from 27 to 270~km/s. 
We find a flat relation between the mean \afe\ within the 1\Rmaj\ ellipses and log(\sgmRe), even if limiting to the massive sample with log(\sgmRe/km s$^{-1}$)$>$1.9. However, the relation becomes positive after we exclude the \mgI\ feature in our fits, which agrees with the results from the previous work with other stellar population models, albeit with relatively large scatter. 
It indicates that the spectral fits with Vazdekis models could give basically the consistent predictions of \afe\ with previous studies when the \mgb\ index is used, but do not work well at the \mgI\ band when their \alf-enhanced version is employed in the metal-rich regime. 
We suggest avoiding this rather wide index, which covers 471\AA, as it might suffer from other effects such as flux-calibration issues. 
For reference, we also measure the stellar population radial gradients within 1\Rmaj\ ellipses. Due to the low resolution of age estimations for old objects and the \mgI\ issue, the uncertainties of these gradients cannot be neglected. 
\end{abstract}




\section{Introduction}
\label{sec:Intro}

Galaxies are the basic elements of cosmic structures. A significant fraction of them, especially in galaxy cluster environments, are early-type galaxies (ETGs). ETGs hardly have any ongoing star formation activities. Their mysterious star formation histories and quenching scenarios are imprinted in their galactic stellar population properties. 

The two most often studied stellar population parameters are age and metallicity (\ZH). In addition to them, the abundance ratios between different elements are also important. 
Particularly, the mean stellar alpha-to-iron abundance ratio (\afe) is an indicator of galactic star formation timescale under the assumption of a universal stellar initial mass function (IMF) and constant supernova properties. 

After a galaxy starts to form stars, in the first $\sim$0.1~Gyr \citep{Maoz_10}, only massive stars evolve and explode as core collapse supernovae, enriching the interstellar medium with relatively high \afe. Afterwards, lower mass stars also evolve and explode as Type~Ia supernovae, which ejects lower \afe\ ejecta into the interstellar medium. From then on, the later generations of stars would have lower \afe, and the mean \afe\ of the entire galaxy decreases with time. Therefore a galaxy with shorter star formation timescale would have a higher \afe\ \citep[e.g.,][]{MatteucciTornambe_87,Thomas_99,FerrerasSilk_03}. 

Over the past decades, three empirical relations are well established between stellar population parameters and galactic gravitational potential wells among massive ETGs. The galactic mean stellar age, \ZH, and \afe\ have positive correlations with the galactic velocity dispersions (\sgm) \citep[e.g.,][]{TMB05,McDermid_15,Liu_16c}. 
When including the low-mass range, though the scatter becomes significantly larger, the slopes of the three relations are also positive \citep[e.g.,][]{Liu_16c,Li_18}. 
On average, the more massive ETGs form stars earlier, are more metal-rich, and have shorter star formation timescale. 
These are important information to constrain the picture of the formation and evolution of ETGs. 

However, while the present-day cosmological simulations can naturally reproduce the age-\sgm\ and \ZH-\sgm\ relations of ETGs \citep[e.g.,][]{Monachesi_19}, their outputs of the \afe-\sgm\ relations do not match the observations. Simulations can only produce the positive slopes after involving special mechanisms, e.g., more top-heavy IMFs, different supernova properties, and AGN feedback \citep[e.g.,][]{Arrigoni_10,Yates_13,Segers_16,Vincenzo_18}. Especially, although the radio mode AGN feedback is a popular recipe in simulations, it may cause difficulties in producing both the positive \ZH-\sgm\ and \afe-\sgm\ relations simultaneously \citep{Okamoto_17}. 

On one hand, there is space for theorists to explore the physical reasons behind such puzzling \afe-\sgm\ relations. On the other hand, from observations, there are limitations in the previous stellar population analysis. 
In general, there are two major approaches for stellar population analysis. One is comparing the spectral line indices from real measurements with the model prediction. For example, some popular stellar population models (e.g., \citealt{TMB03,GS08,TMJ11}) are based on the widely used Lick index system \citep{W94}. 
The other one is fitting the galactic spectra over a wide wavelength range by model spectral templates. Some representative fitting programmes are \ppxf\ \citep{CE04,Cappellari_17}, \texttt{STARLIGHT} \citep{Fernandes_05}, \texttt{FIREFLY} \citep{Wilkinson_15,Wilkinson_17}, and \texttt{alf} \citep{Conroy_18}. 
While the former method only makes use of the information from the prominent features and the requirement of signal-to-noise ratio (S/N) is relatively low, the latter one takes into account the information over the entire spectra and requires a higher S/N. 

Although the spectral fitting methods are widely used these years, they have been mostly limited to the measurements of age and metallicity so far. 
\texttt{alf} can estimate the abundance of 18 individual elements by fitting the corresponding model spectral with the MCMC algorithm. However, it is time consuming and does not provide the measurements of integral \afe\ straightforwardly. 
A model that allows the estimation of integral \afe\ from a full spectral fitting approach is \citet[][hereafter V15]{V15}. It provides the spectral templates with the variation of \afe. 
Furthermore, while most other popular models are constructed on solar \alf\ abundance stellar isochrones, V15 employed \alf-enhanced isochrones. 
Therefore, in this work, we aim to study the \afe\ of ETGs by means of the spectral fitting technique with V15 models. 

Besides the global properties, the radial gradients of the stellar population parameters of ETGs are also meaningful. 
The flatness of these radial profiles provides the information of galactic merger histories \citep{Ogando_05}, and the slopes help us to constrain the picture of their quenching scenarios. 
From observations, the measurements of stellar population gradients within 1\re\ of ETGs are basically in agreement \citep[e.g.,][]{SanchezBlazquez_07,Kuntschner_10,OlivaAltamirano_15,Greene_15,GonzalezDelgado_15}. ETGs generally have zero or very mild positive gradients in age and \afe. Their \ZH\ gradients are negative at high masses and become shallower with the decrease of mass. 
Simulations also support these results \citep[e.g.,][]{Tortora_11}. However, due to the limitations in previous \afe\ measurements that we mentioned above, we also estimate the \afe\ gradients of our sample ETGs using the spectral fitting method in this work. 

Note that there is a caveat for the full spectral fitting method. \citet{Ge_18} measured the mock single stellar population spectra with different signal-to-noise ratio (S/N) level by \ppxf\ and studied the bias of age and metallicity measurements as functions of S/N. They found that the age and metallicity estimations could be 20\% older and 10\% lower at S/N = 10. The bias decreases with S/N, and could be neglected at S/N about 60. 
Therefore, to minimise the bias, we only adopt high S/N data in this study. 

MaNGA survey \citep{Bundy_15} is an integral field unit (IFU) survey of nearby galaxies (redshift from 0.01 to 0.15) with a wavelength coverage over 3600-10300\AA, using a 2.5-meter telescope at Apache Point Observatory. 
In this work, we study the mean \afe\ and \afe\ gradients of 196 MaNGA ETGs with high S/N. 

The paper is structured as follows: 
We introduce our data and sample in Section~\ref{sec:DataSample} and describe our stellar population analysis in Section~\ref{sec:Analysis}. We test our methods and sample representativeness in Section~\ref{sec:Test}. We present our results in Section~\ref{sec:Result} and compare different methods of stellar population analysis in Section~\ref{sec:Discussion}. Our conclusions are listed in Section~\ref{sec:Conclusion}.


\section{Data and Sample}
\label{sec:DataSample}

\subsection{Data}
\label{sec:data}

The data for our analysis in this work are from the SDSS Data Release 15 (DR15), MaNGA Product Launch 7 (MPL7). 
All raw MaNGA data are firstly reduced by the Data Reduction Pipeline \citep[DRP;][]{Law_16}. Then these science-ready data cubes are analysed by the Data Analysis Pipeline \citep[DAP;][]{Westfall_19}. In this step, DAP spatially bins the data cubes in a certain way, and measure the physical parameters from the stacked spectrum in each bin. 
Our work is based on the DAP products. The binning method we adopted is the Voronoi binning \citep{CC03} at a target S/N of 10. 

DAP produces both the three dimensional LOGCUBE files and the two dimensional MAPS files. 
The LOGCUBE files are similar to the DRP cubes. Behind each spaxel in LOGCUBE files, the spectra in different dimensions correspond to either the spectrum stacked from the bin it belongs to or the fitting outputs from this stacked spectrum. 
The MAPS files record a range of parameters that are measured from the corresponding spectra in DAP cubes. 

\subsection{Sample}
\label{sec:sample}

\citet[][hereafter L18]{Li_18} estimated the velocity dispersions within 1\re\ (\sgmRe) of MaNGA MPL5 galaxies by dynamical modelling. We adopt their \sgmRe\ measurements and take the MPL5 galaxies as our parent sample. 
L18 also provided the morphological classification of these galaxies, which was based on the {\it GalaxyZoo} project \citep{Lintott_08,Lintott_11}. We adopt this traditional classification to select out ETGs, so that our results are comparable to most previous studies in this field \citep{Cappellari_16}. 

For the purpose of studying high-S/N spectra, we only select the ETGs with the median S/N within 1\re\ higher than 30. The median S/N within 1\re\ is provided by DAP MAPS files. This criterion ensures the S/N of our binned spectra all above 50 and mostly higher than 80 (See Figure~\ref{fig:histSNR} and further description in \S\ref{sec:bins}), high enough for a relatively accurate spectral fitting analysis (see \S\ref{sec:Test:mockSpec}). 
Because we check the representativeness of our sample by means of the Lick index analysis (\S\ref{sec:Test:LickI}), our sample only includes the galaxies with all the related indices (Fe4383, \mgb, Fe5270, Fe5335) available. These galaxies also have useable H$_\beta$ index for age estimations. 
Under these criteria, our sample contains 196 ETGs. Their \sgmRe\ ranges from 27 to 270~km/s. 

\begin{figure}
 \includegraphics[width=\columnwidth]{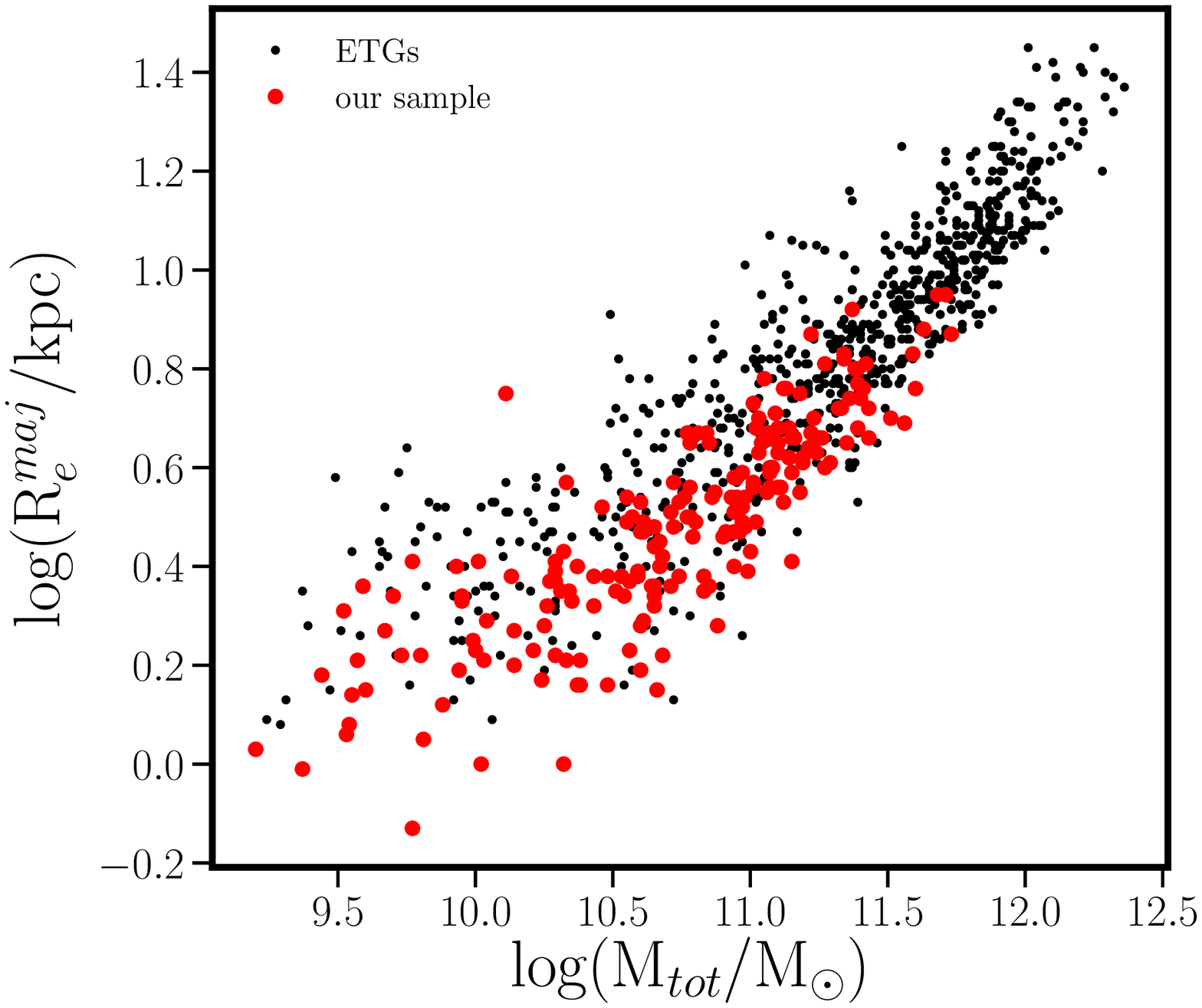}
 \caption{
 The mass-size diagram of our sample (red circles) and the ETGs from MPL5 (black dots). The masses are two times of the dynamical masses that are enclosed in 1\re\ spheres. The sizes are the major axis of the half-light isophote. Both the masses and sizes are adopted from L18.}
 \label{fig:MRDsamp}
\end{figure}

Figure~\ref{fig:MRDsamp} shows the mass-size distribution of our sample galaxies (red circles) comparing to the ETGs from the entire parent sample (black dots). The mass and size measurements are from L18. The masses are two times of the dynamical masses that are enclosed in 1\re\ spheres. The sizes are the major axis of the half-light isophote (\Rmaj). 

Note that MaNGA applies a volume limited target selection. The more massive galaxies are from higher redshifts hence have lower S/N. Therefore our high-S/N selection criterion excludes the galaxies that are more massive than 10$^{11.8}$\Msun. 
According to the studies from the literature, the global stellar population properties of the most massive and intermediate-mass ETGs have the same trends with galactic velocity dispersions \citep[e.g.,][]{TMB05,McDermid_15}. Thus the lack of ETGs at massive end should not affect our major results, at least for the relations between global stellar population parameters and \sgm. 

\section{Data Analysis}
\label{sec:Analysis}

\subsection{Spatial Resolving}
\label{sec:bins}

\begin{figure}
 \includegraphics[width=\columnwidth]{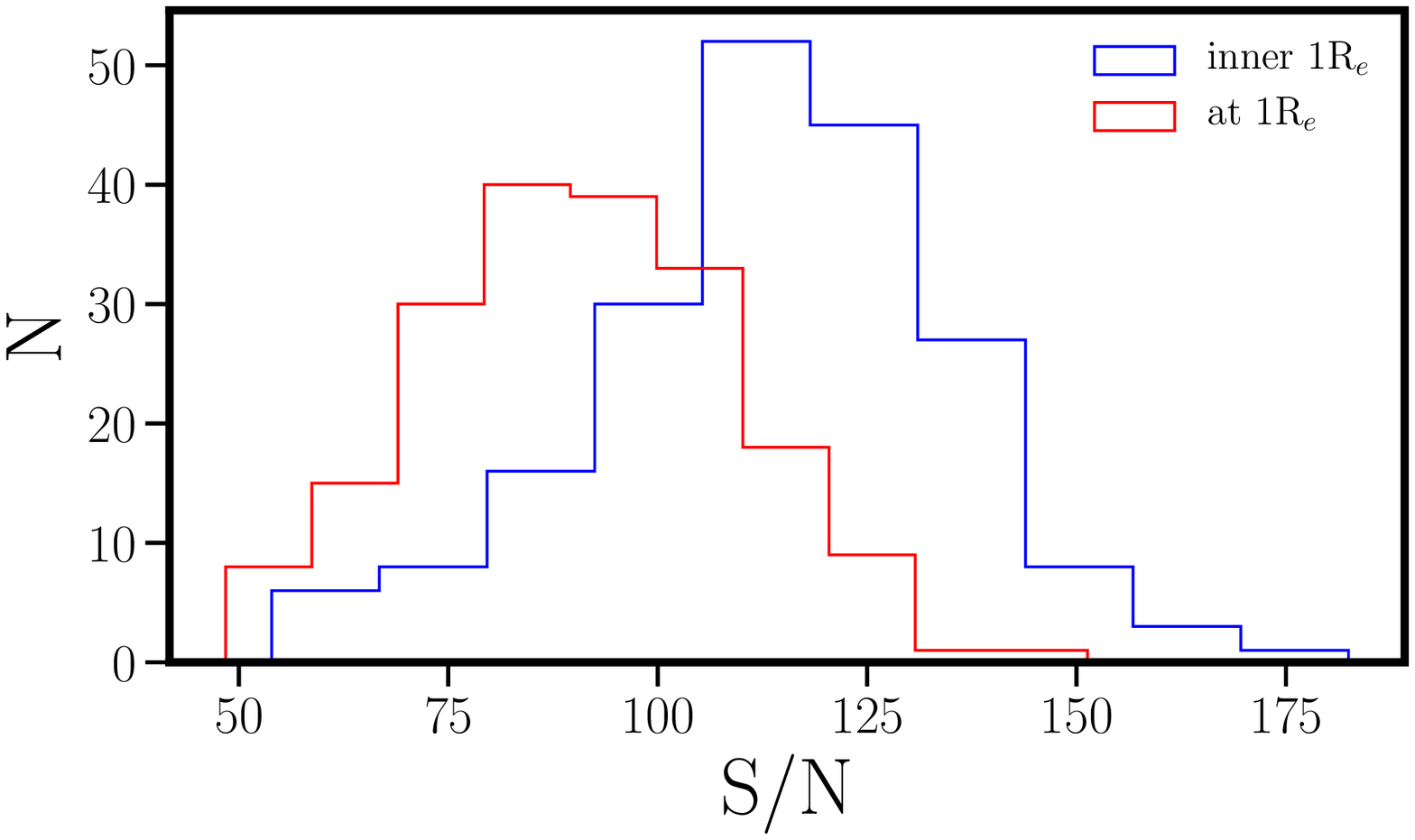}
 \caption{
 The histograms of the S/N of our stacked spectra. Blue and red colours indicate the spectra stacked from the inner 1\Rmaj\ ellipse and the radial bin at \Rmaj, respectively. The S/N is measured from the full spectral fits.}
 \label{fig:histSNR}
\end{figure}

We combine the light from the inner 1\Rmaj\ ellipse of our sample ETGs for the study of their global stellar population properties. The geometry calculation is based on the photometric position angle and ellipticity provided by \citet{Graham_18}, together with the \Rmaj\ from L18. The blue histogram in Figure~\ref{fig:histSNR} illustrates the S/N distribution of these stacked spectra. The S/N shown in this figure is calculated by the best fitting spectra and the residuals. All the S/N is higher than 50, enough for our spectral fitting analysis (see \S\ref{sec:Test:mockSpec}). 

When studying the stellar population gradients, we equally divide the galactic major axis in log space with the interval of 0.08~log(\Rmaj), starting from 0.4~\Rmaj. The central 0.4~\Rmaj\ ellipses are the innermost bins. Then we stack the spectra within each elliptical annulus. 

We measure the one-dimension radial gradients within 1\Rmaj\ along the major axis. 
From Figure~\ref{fig:histSNR}, the S/N of the radial bins at 1\Rmaj\ (red histogram) is acceptable for spectral fits (see \S\ref{sec:Test:mockSpec}). 
Beyond 1\Rmaj, however, the S/N of the radial bins could be lower than 10, insufficient for an accurate spectral fitting analysis. Therefore we only calculate the gradients within 1\Rmaj. 

For the gradient calculations, we find the r-band luminosity weighted mean radius of each radial bin and fit the linear relation between log(R$_{mean}$/\Rmaj) and \ZH\ or \afe. We take the fitted slopes as radial gradients. 

\subsection{Full spectral fits}
\label{sec:fullSpecFit}

For each galaxy, we calculate the median S/N between 4800 and 5500~\AA\ of the spectrum behind every spaxel in the DAP LOGCUB files. The signal and noise we adopt for this calculation are simply the flux and error provided by DAP. When doing stacking we only combine the spectra with S/N higher than 5. 
The DAP files provide the precise fits of emission lines \citep{Westfall_19}. 
We subtract the emission components before the fits. 

For each stacked spectrum with emission line subtraction, we apply the \ppxf\ spectral fitting programme to fit the r-band luminosity weighted stellar population parameters. The templates are made from the V15 models which are constructed on the BaSTI scaled-solar and \alf-enhanced isochrones with a Salpeter IMF. \ppxf\ produces the best fitted spectra by weighting templates. 

The templates with \afe\ = 0.0 and 0.4 correspond to the models with BaSTI scaled-solar and \alf-enhanced isochrones respectively. At either \afe, V15 provides the templates with ages ranging from 0.03 to 14~Gyr with narrow steps and \ZH\ = [-2.27, -1.79, -1.49, -1.26, -0.96, -0.66, -0.35, -0.25, 0.06, 0.15, 0.26, 0.40]. 
We do not use these original templates for our fits. In \ZH\ dimension, we only choose the templates at \ZH\ = -1.49, -1.26, -0.96, -0.66, -0.35, 0.06, 0.26, and 0.40, in order to have similar intervals across the full range. At each \ZH\ and \afe, we linearly interpolate the templates in log(age) space from log(0.06~Gyr) to log(14~Gyr) to obtain 24 ages with equal interval. 

V15 showed examples of the spectral fits of galactic spectra. They only fitted the spectra from 4800\AA\ to 5500\AA. 
In this work, we adopt the same wavelength range for our fits. 
The fit of each galactic spectrum iterates twice. In the first step, we apply 10-order both additive and multiplicative polynomials in order to obtain fine fits for bad pixel clipping. 
In the second step, we take the kinematic estimations from the first step as the initial guess, assume a reddening curve given by \citet{Calzetti_00} and do not apply any polynomials. The initial guess of E(B-V) is the value provided by the DAP map. 
In both steps, the templates are normalised by the r-band luminosity. 
The wavelength resolution of V15 templates is 2.5~\AA. Although it is similar to that of the MaNGA spectra, during the fits we broaden it to the accurate MaNGA spectral resolution, which is a function of wavelength. 

\subsection{Lick Index Analysis}
\label{sec:3I}

Before applying our spectral fits for the main results, we did test work with Lick index analysis (see \S\ref{sec:Test:LickI}). In our test work, the index analysis relates to two stellar population models, \citet[][hereafter TMJ11]{TMJ11} and V15. 

TMJ11 is constructed with a Salpeter IMF. It has the same spectral resolution as V15. TMJ11 provides the model grids of Lick indices that are produced from two sets of stellar evolutionary tracks. One is entirely based on \citet[][hereafter C97]{C97}, while the other is based on C97 at low metallicity but on \citet[][hereafter P00]{P00} at \ZH$\geqslant-$0.33. 
TMJ11 grids have age = [0.1, 0.2, 0.4, 0.6, 0.8, 1, 2, 3, 4, 5, 6, 7, 8, 9, 10, 11, 12, 13, 14, 15]~Gyr, \ZH\ = [-2.25, -1.35, -0.33, 0.0, 0.35, 0.67], and \afe\ = [-0.3, 0.0, 0.3, 0.5]. 
We keep their age dimensions and linearly interpolate the \ZH\ and \afe\ dimensions across their full ranges with a step of 0.05 for our analysis. 

For V15, we construct the model grids by measuring the Lick indices of their model templates. Still based on the V15 model templates that are constructed on BaSTI scaled-solar and \alf-enhanced isochrones with a Salpeter IMF, we firstly measure the Lick indices of all the original spectral templates. Then, similar to TMJ11, we keep their age dimension and linearly interpolate the \ZH\ and \afe\ dimensions across their full ranges with a step of 0.05. 

DAP MAPS files provide the Lick index measurements of the spectrum behind each Voronoi binned spaxel at a 2.5~\AA\ wavelength resolution, which is the same as the resolutions of TMJ11 and V15 model spectra. We calculate the mean indices by weighting the index maps within the stacking areas by the r-band luminosity. 
Then we choose certain indices to fit the three stellar population parameters by searching the minimum reduced $\chi^2$ solutions from the interpolated TMJ11 or V15 grids.


\section{Test Work}
\label{sec:Test} 

Before our proposed scientific analysis, we do the test work for three major purposes: (1) Check if our spectral fitting programmes work well; (2) Figure out which kinds of data quality, especially in terms of S/N, allow a reliable full spectral fitting analysis; (3) Ensure our high-S/N subsample is representative. 

\subsection{Mock Spectra}
\label{sec:Test:mockSpec}

\begin{figure*}
 \includegraphics[width=0.5\columnwidth]{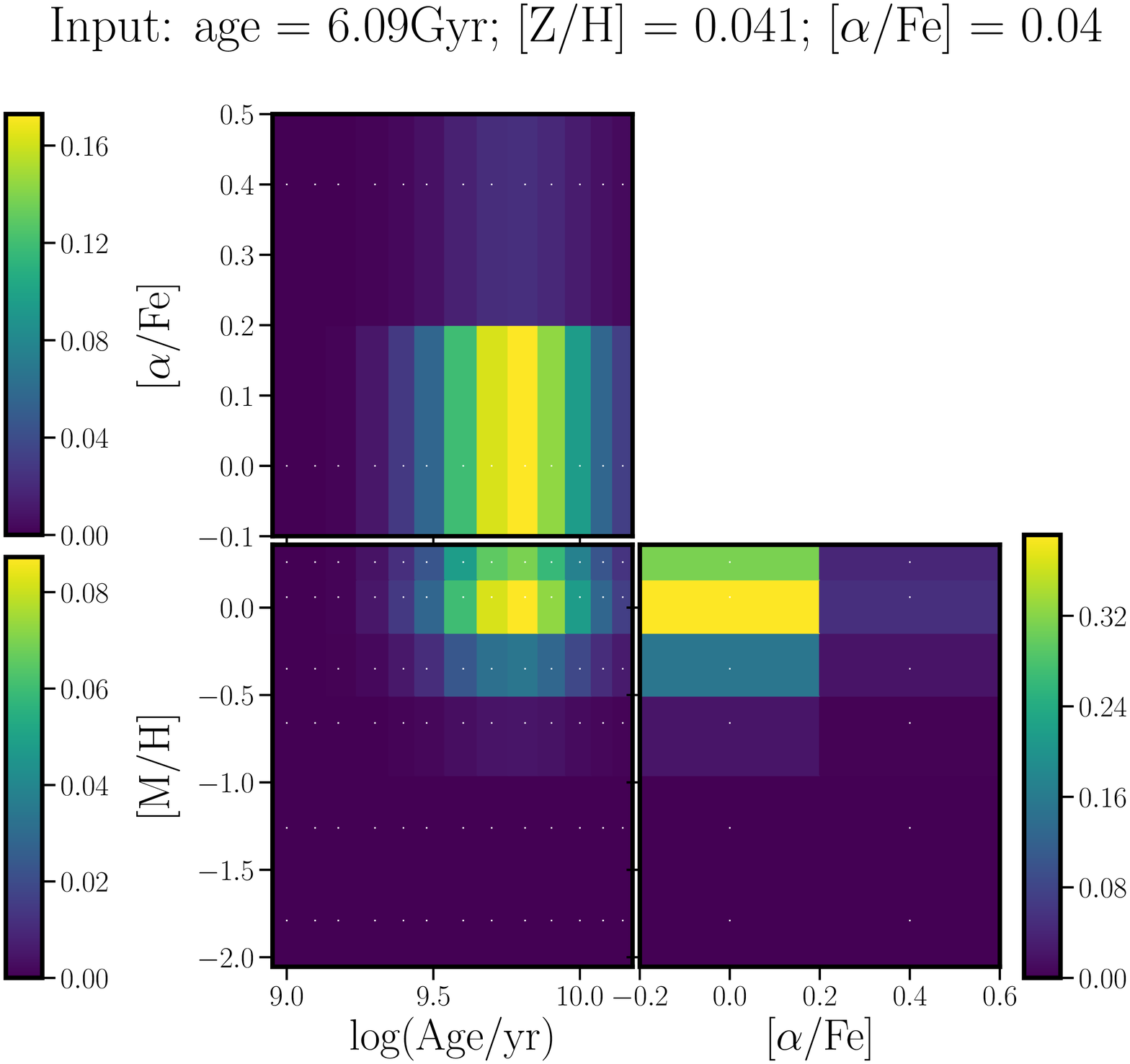}
 \includegraphics[width=0.5\columnwidth]{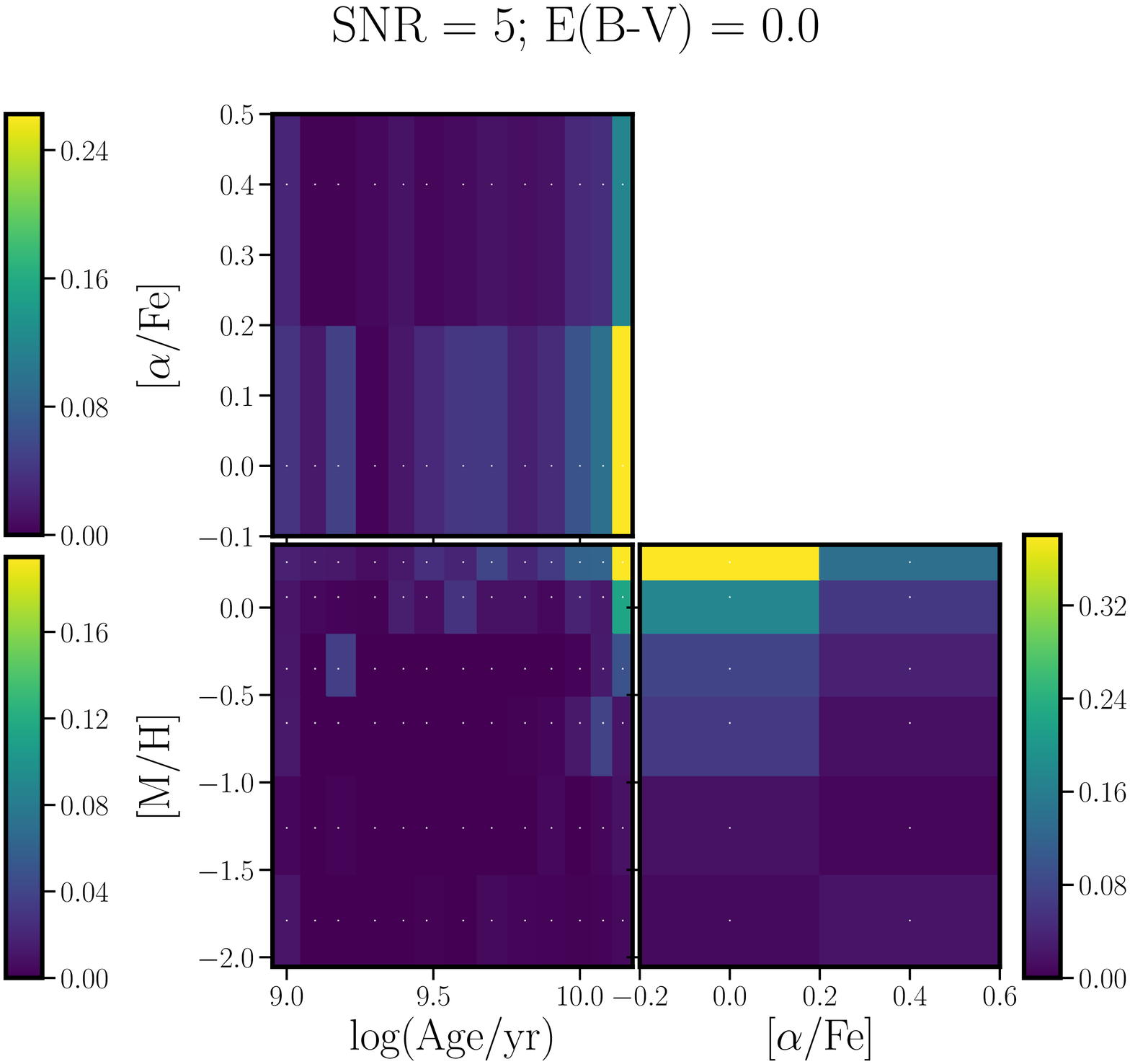}
 \includegraphics[width=0.5\columnwidth]{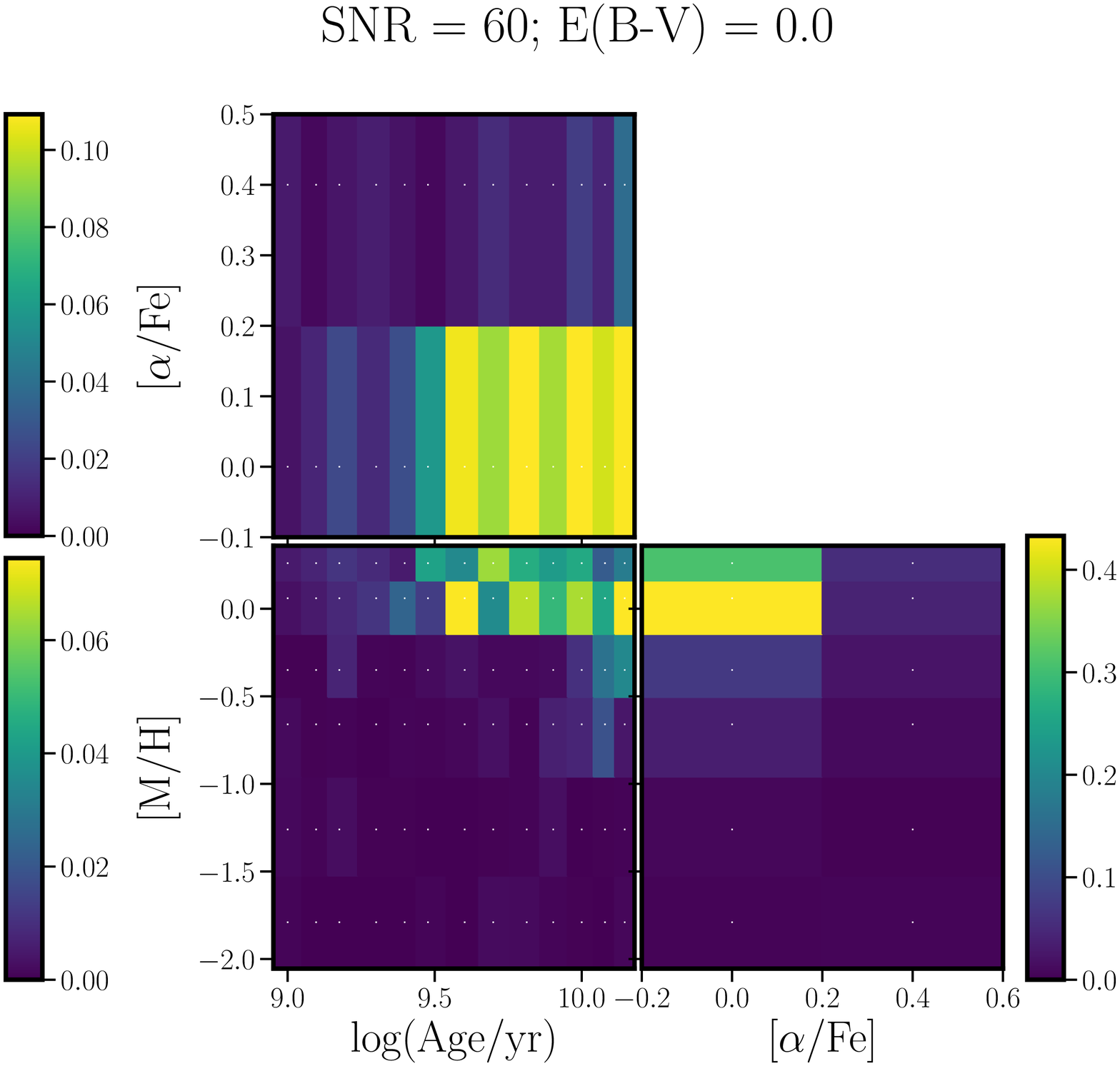}
 \includegraphics[width=0.5\columnwidth]{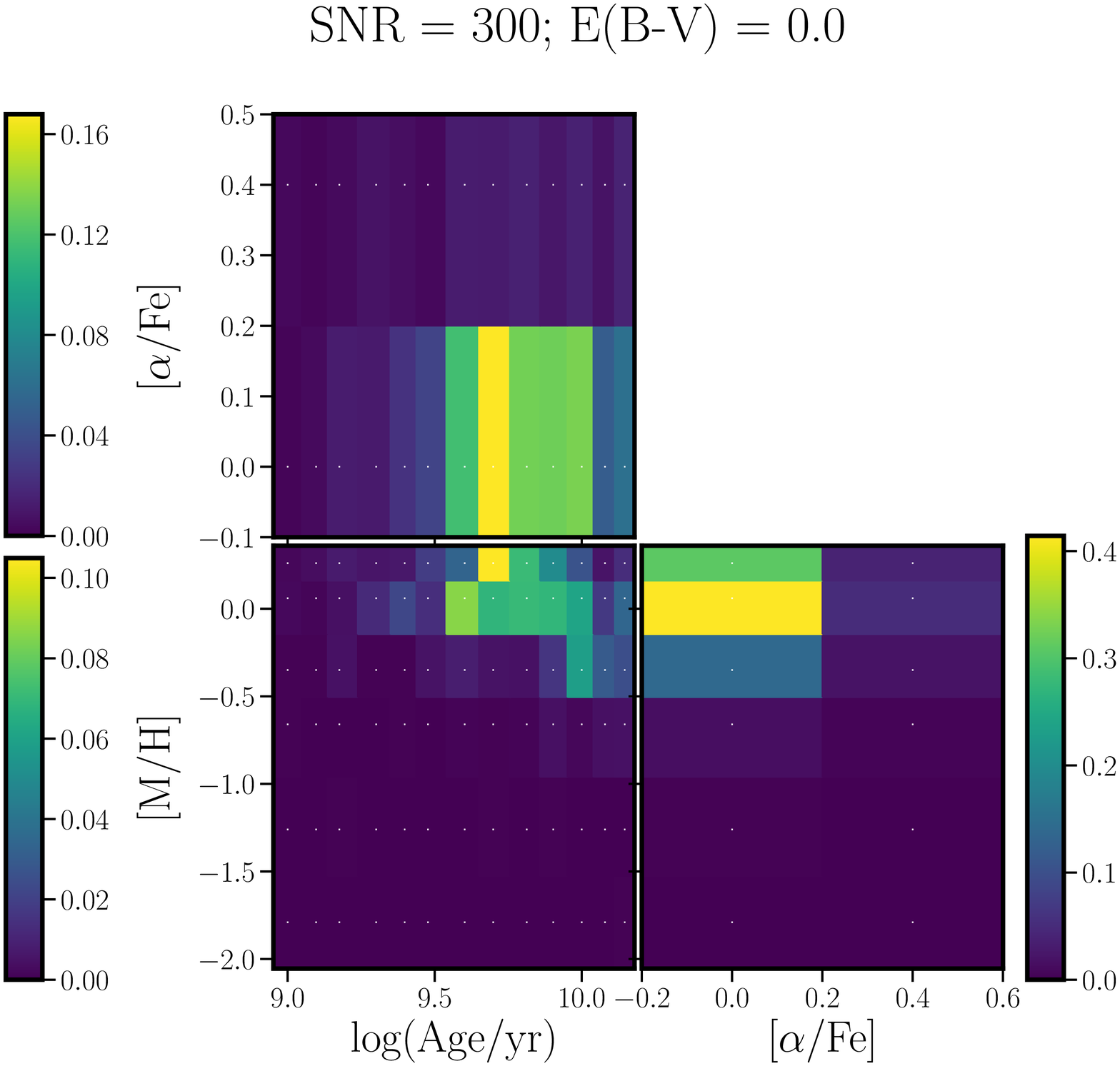}
 
 \includegraphics[width=0.5\columnwidth]{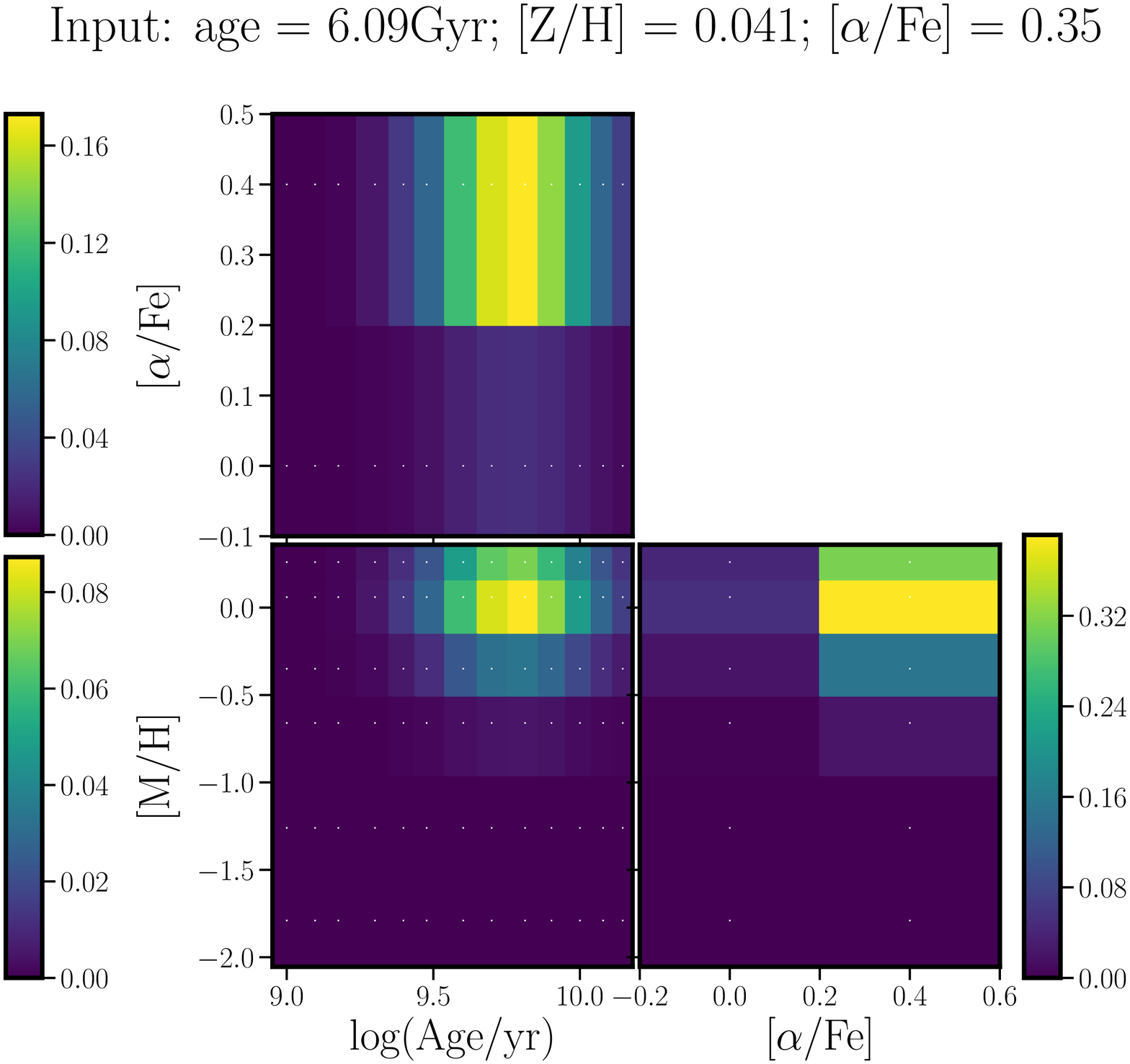}
 \includegraphics[width=0.5\columnwidth]{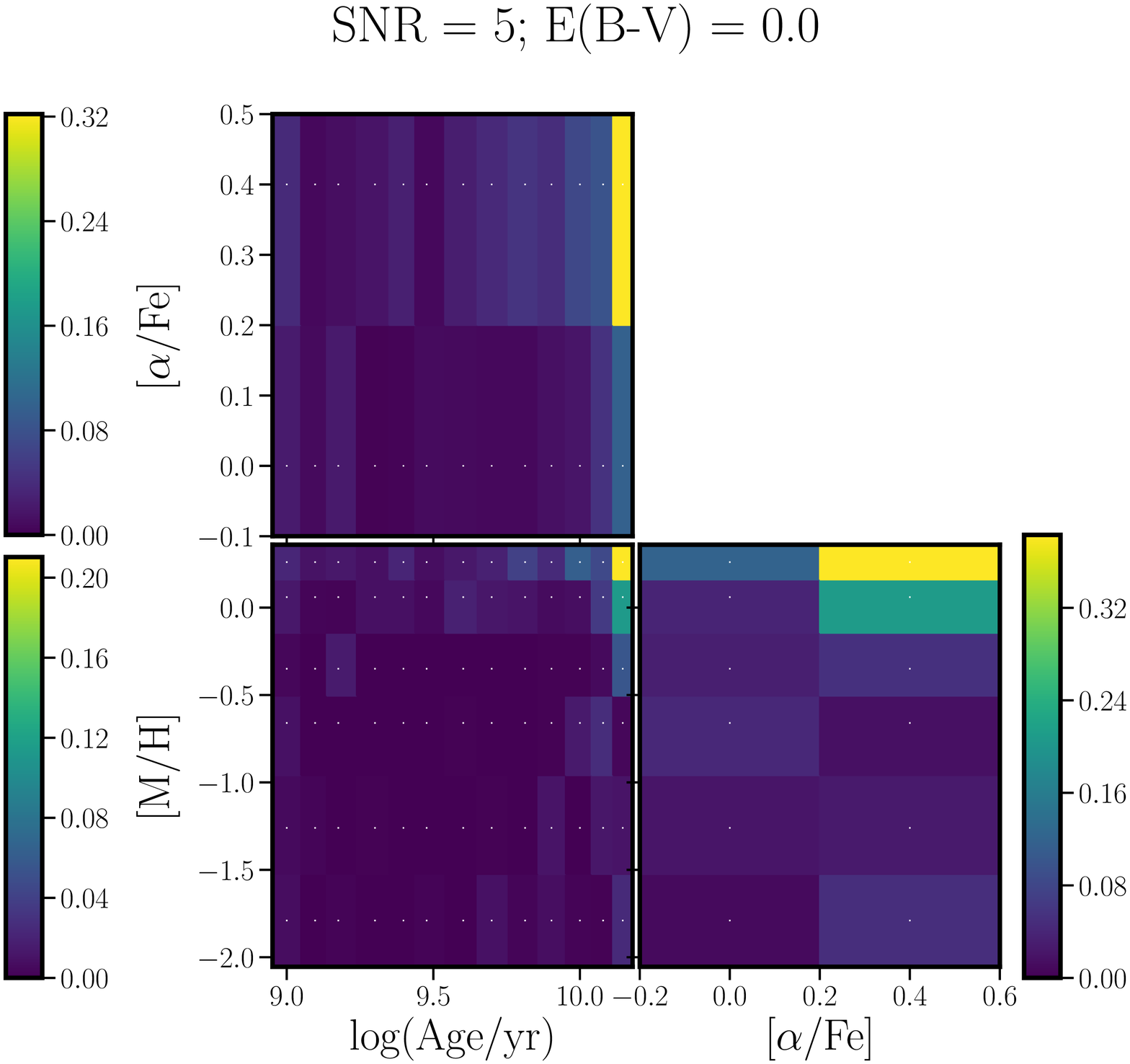}
 \includegraphics[width=0.5\columnwidth]{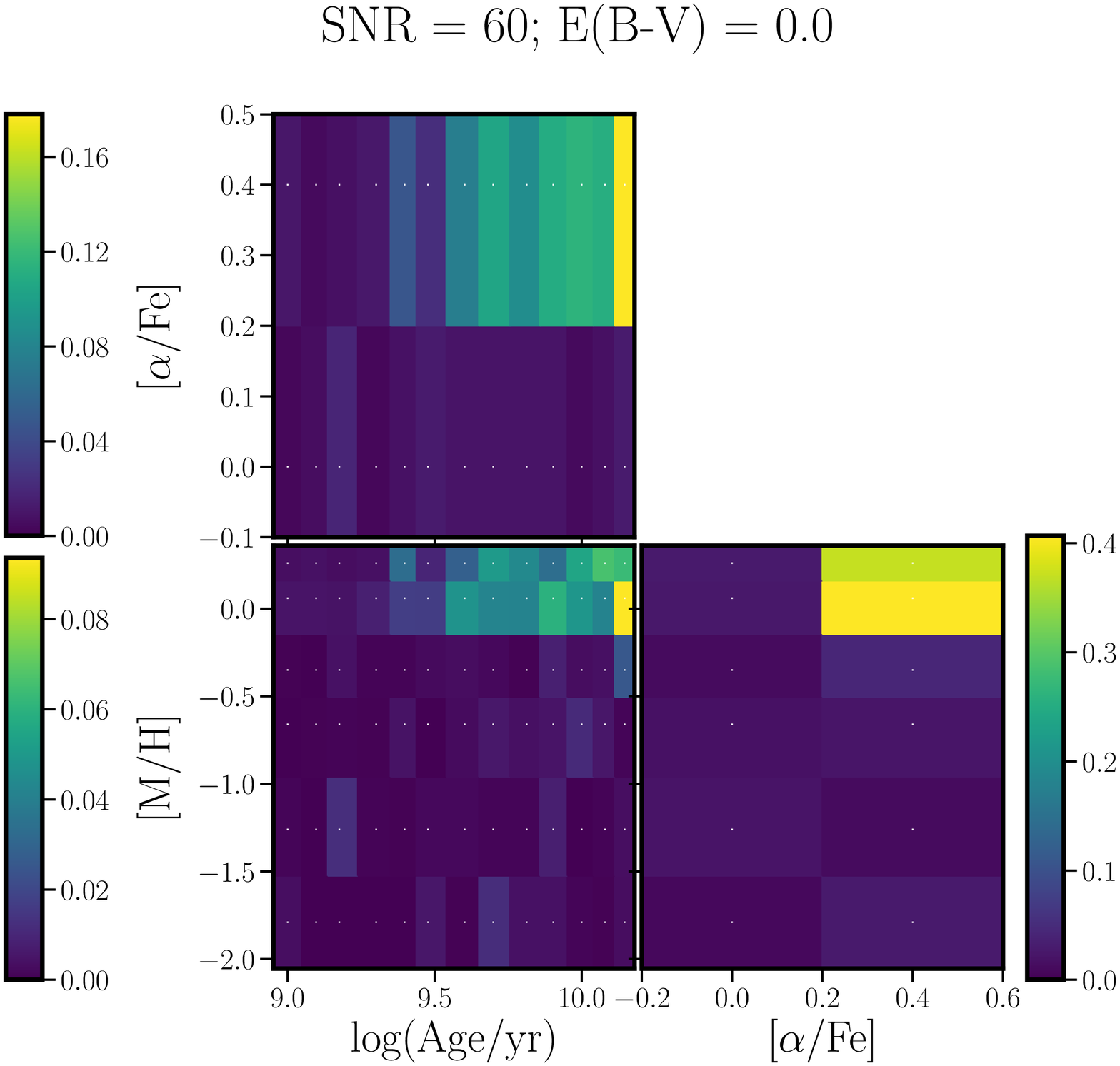}
 \includegraphics[width=0.5\columnwidth]{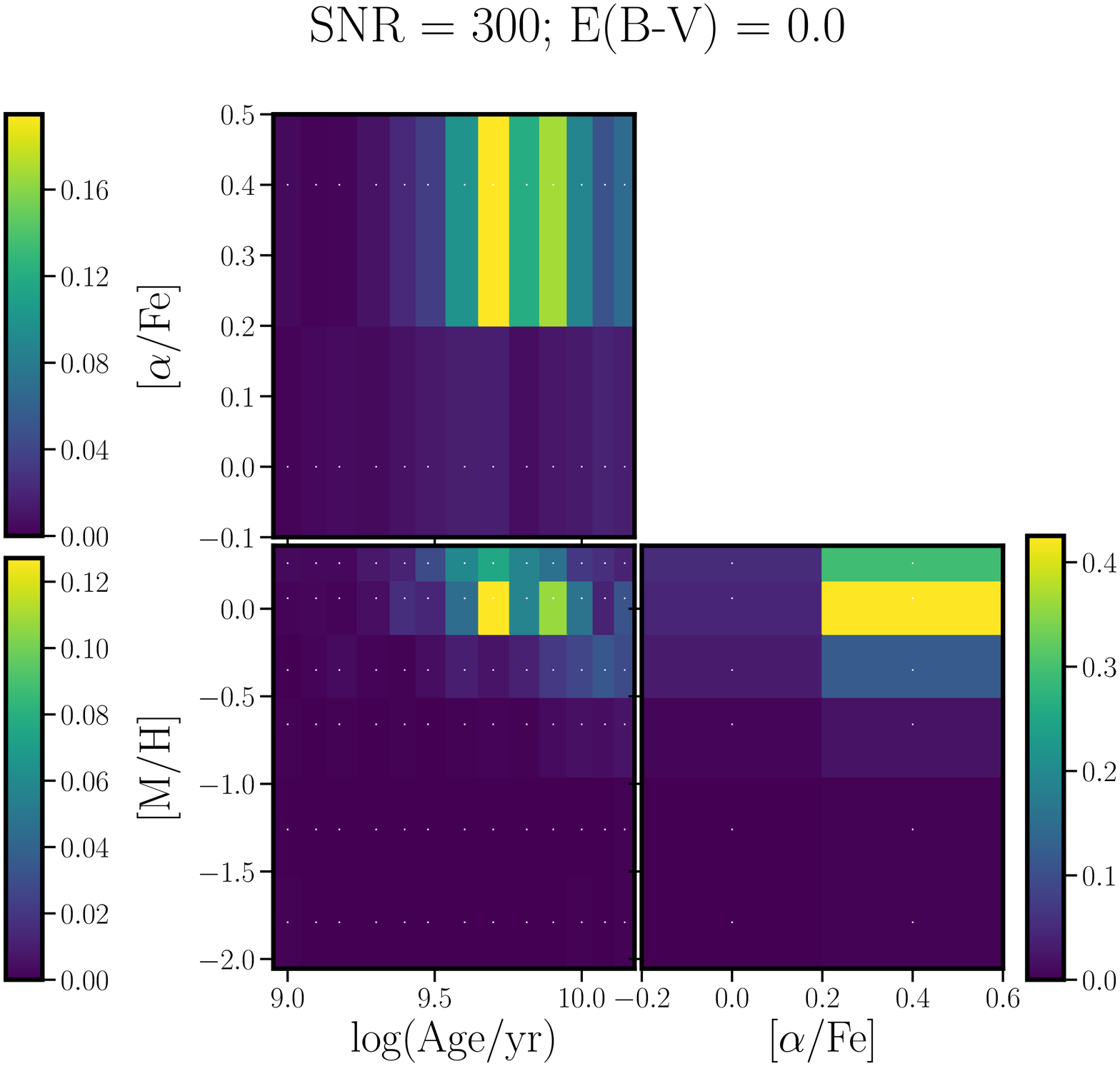}
 
 \includegraphics[width=0.5\columnwidth]{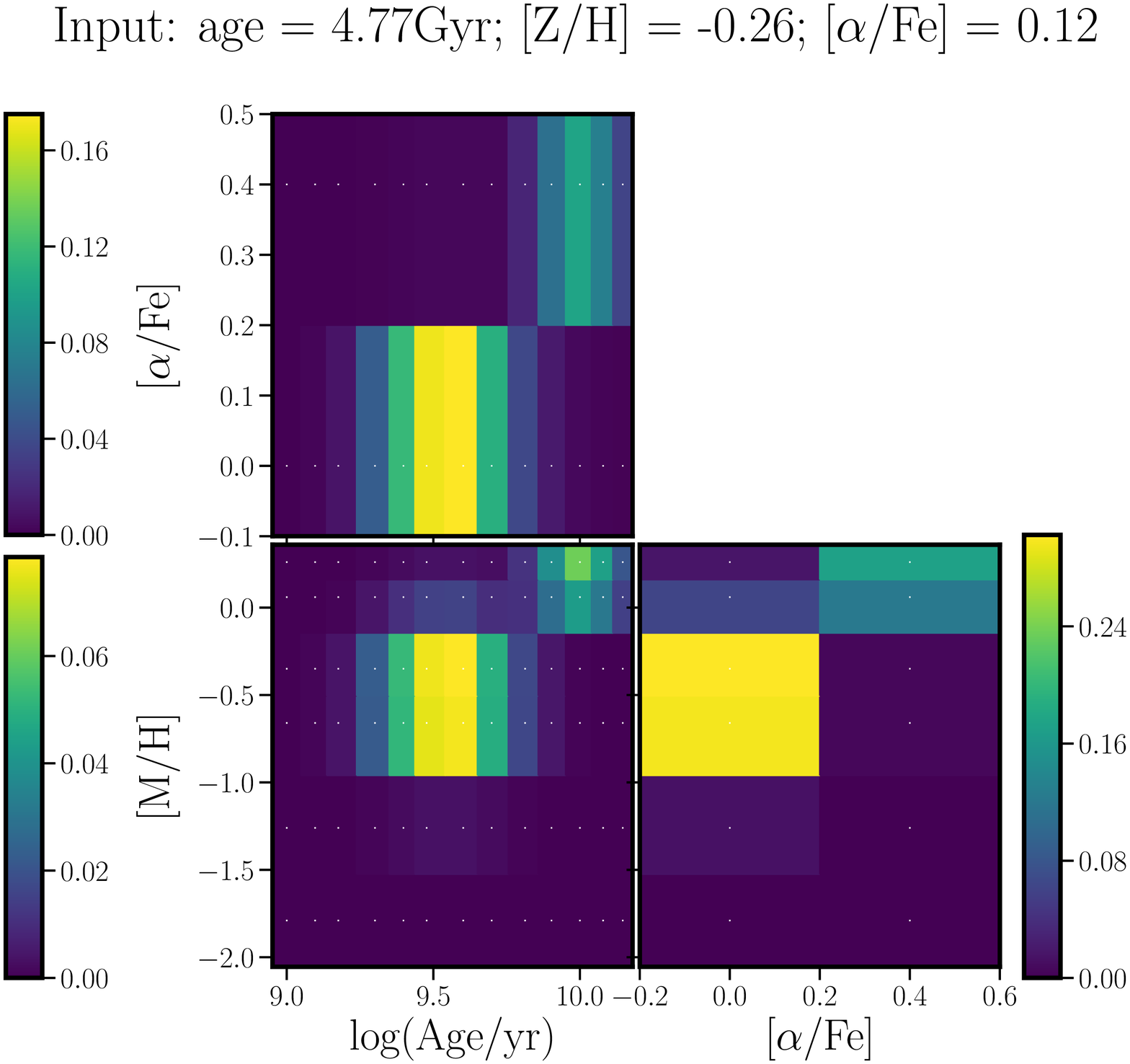}
 \includegraphics[width=0.5\columnwidth]{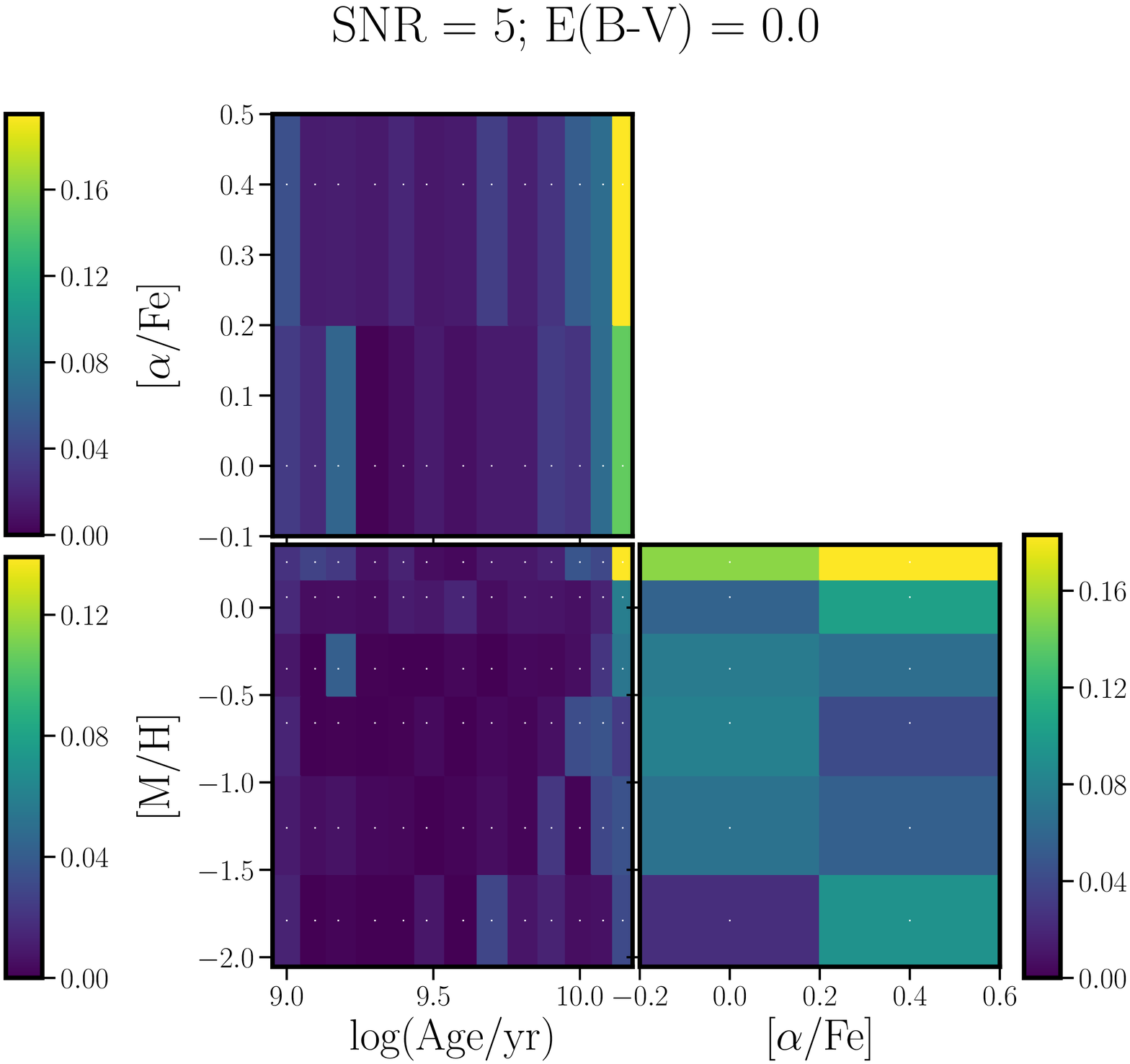}
 \includegraphics[width=0.5\columnwidth]{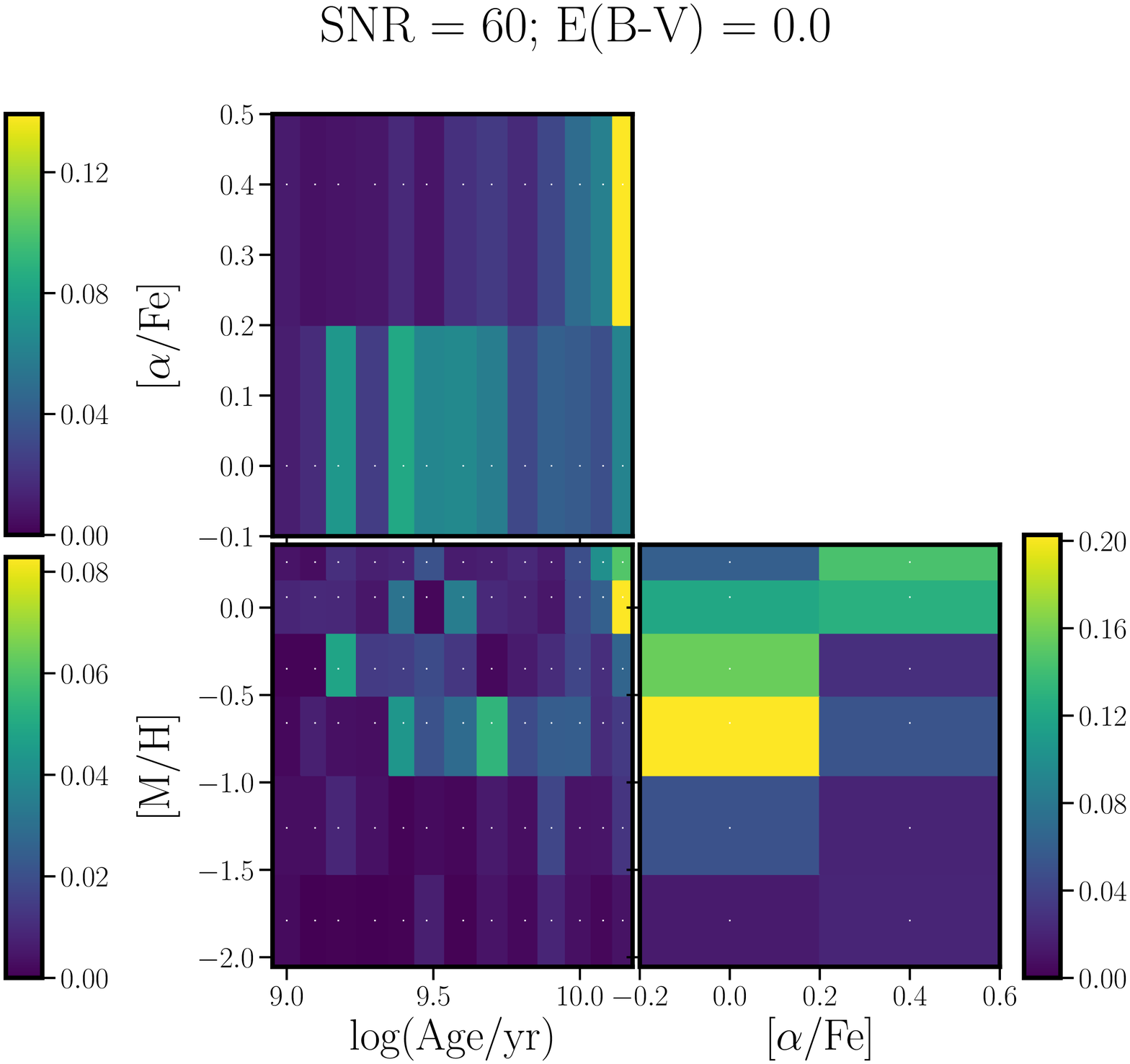}
 \includegraphics[width=0.5\columnwidth]{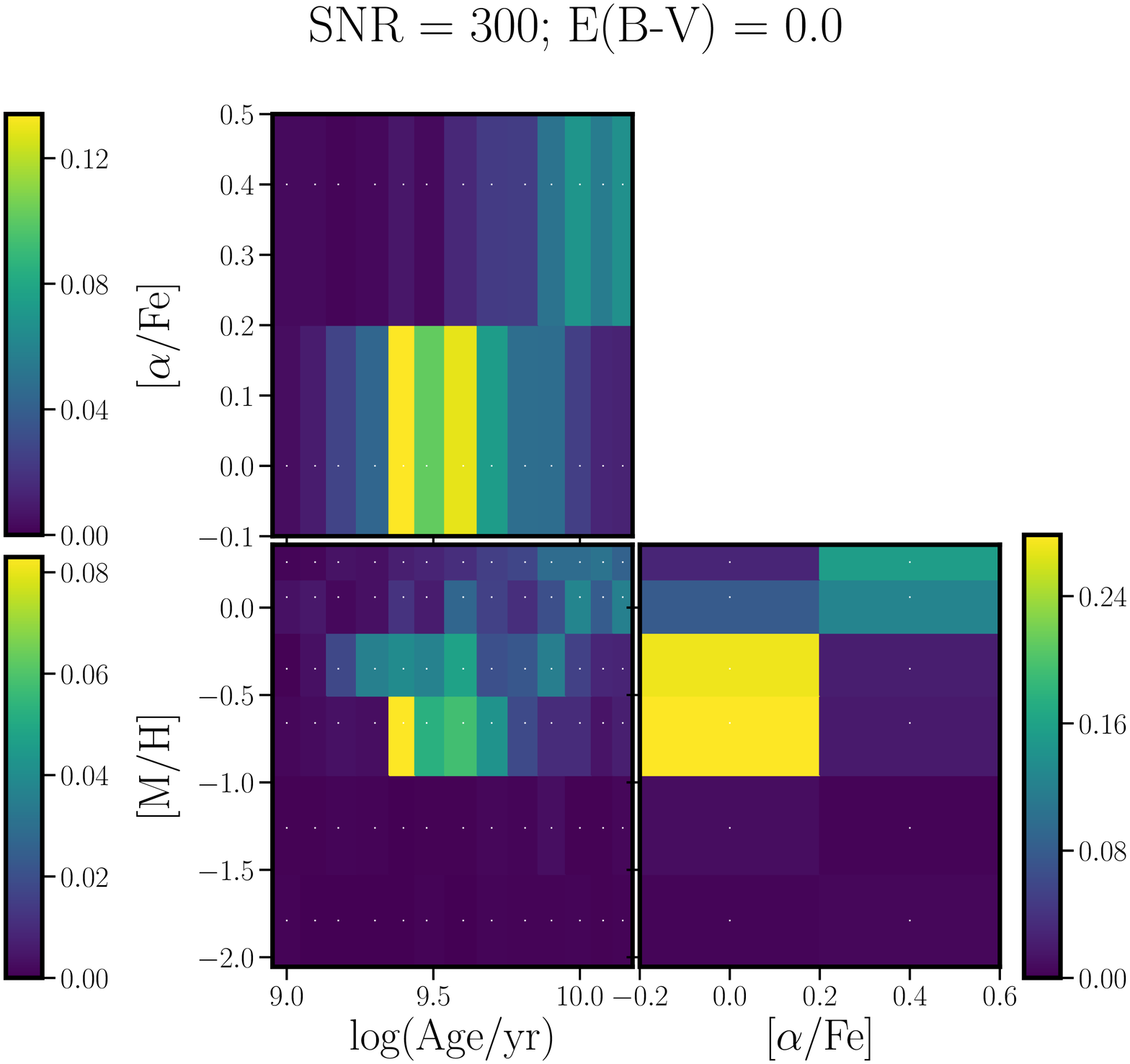}
 
 \includegraphics[width=0.5\columnwidth]{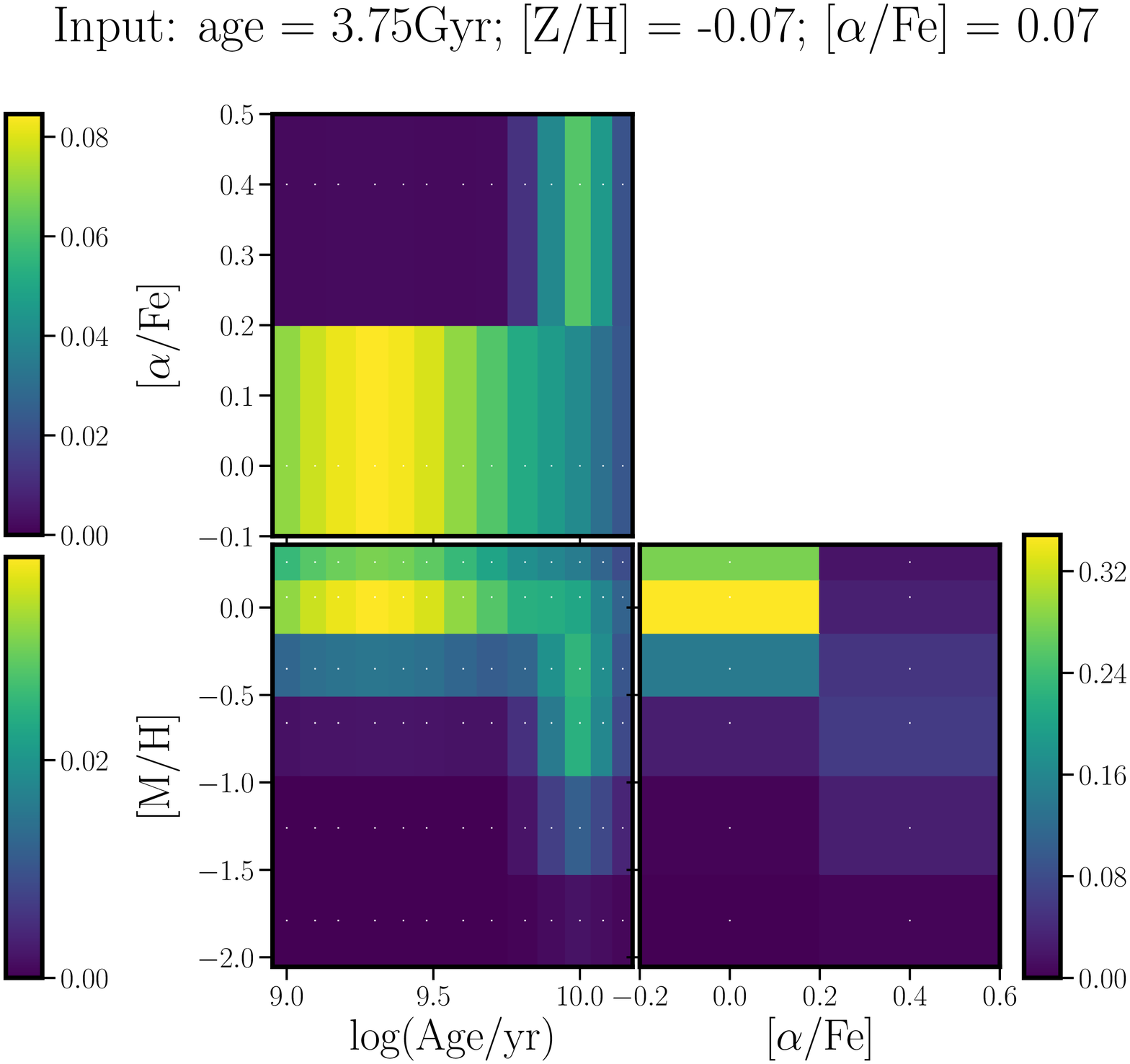}
 \includegraphics[width=0.5\columnwidth]{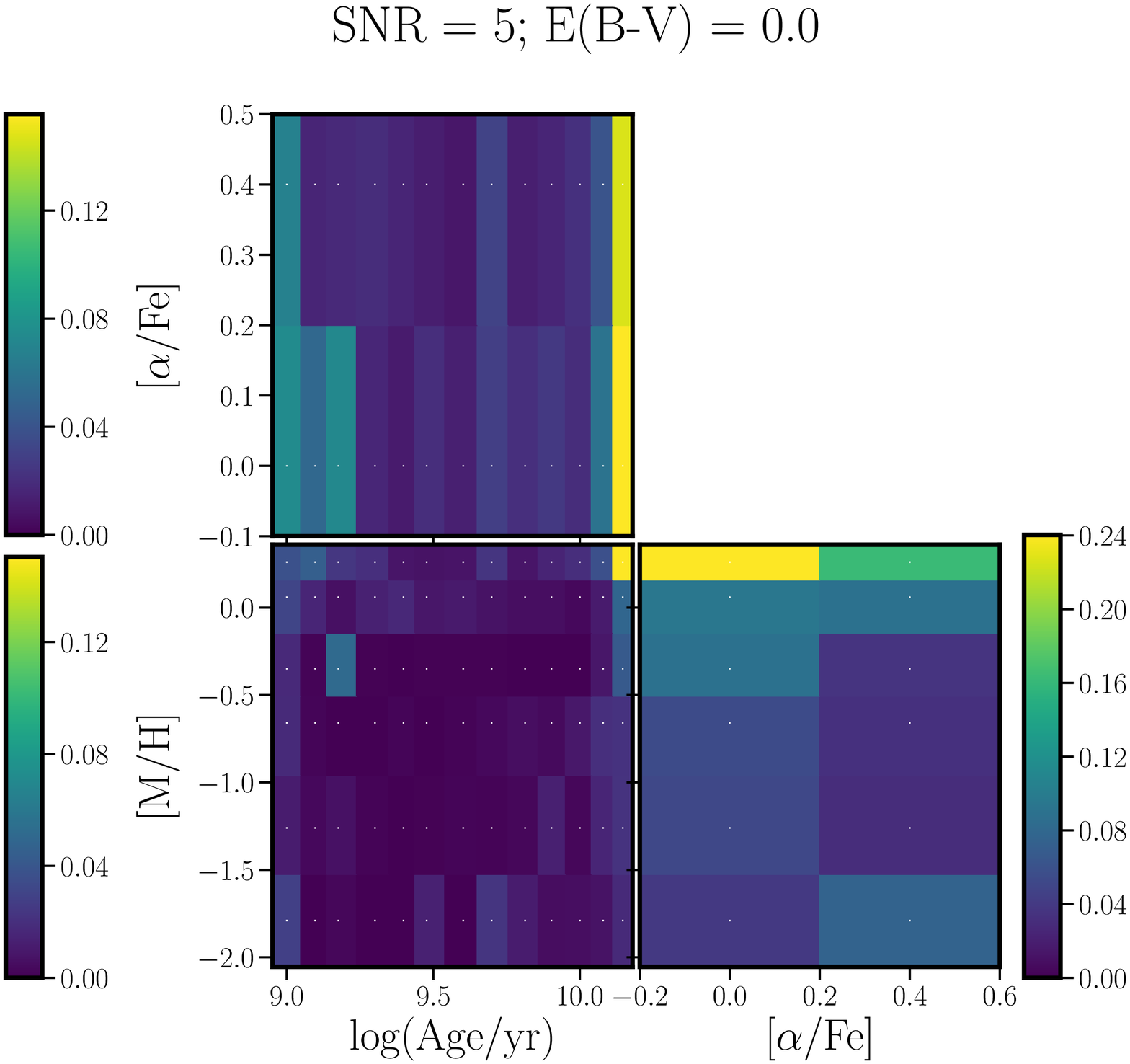}
 \includegraphics[width=0.5\columnwidth]{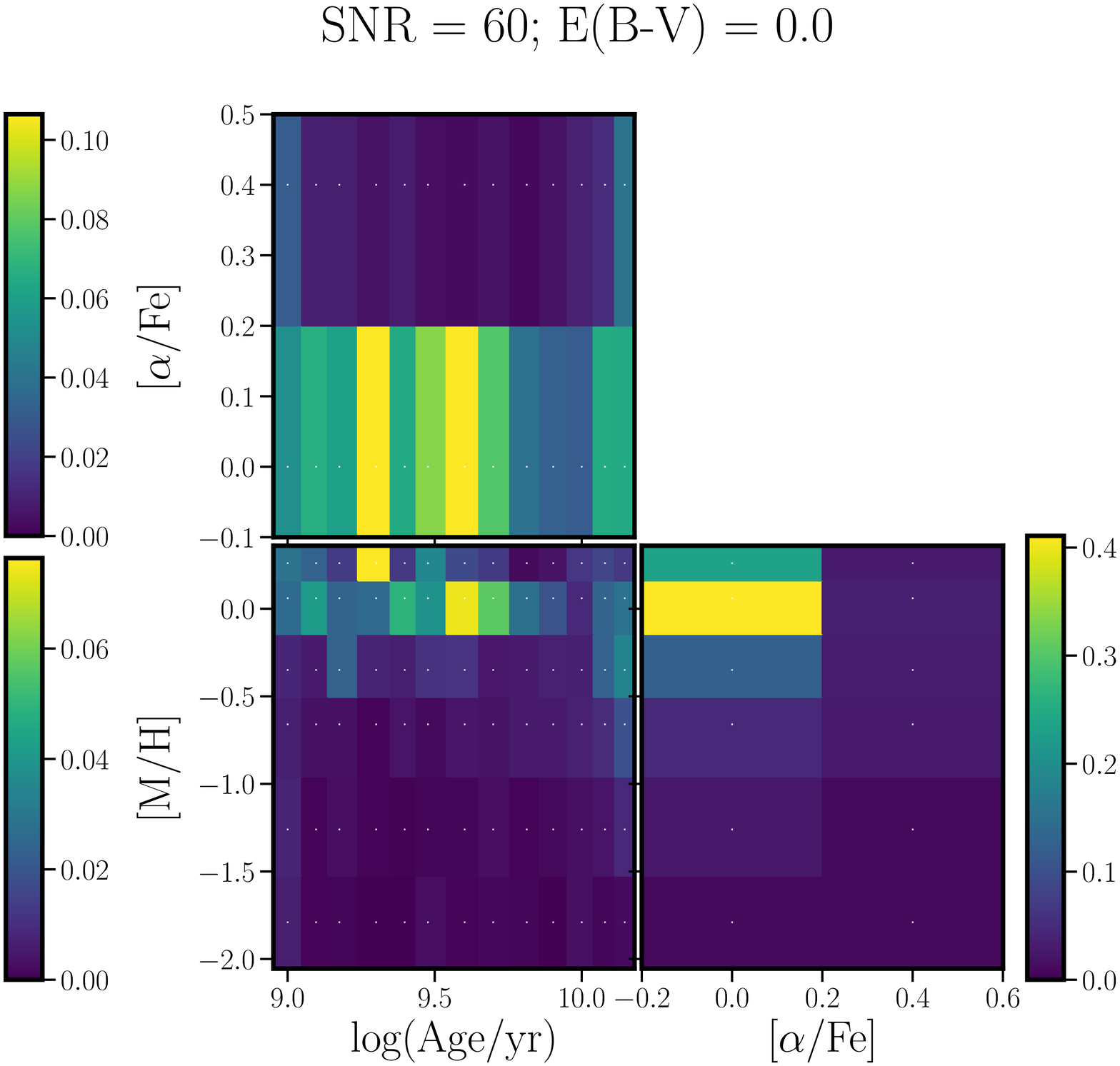}
 \includegraphics[width=0.5\columnwidth]{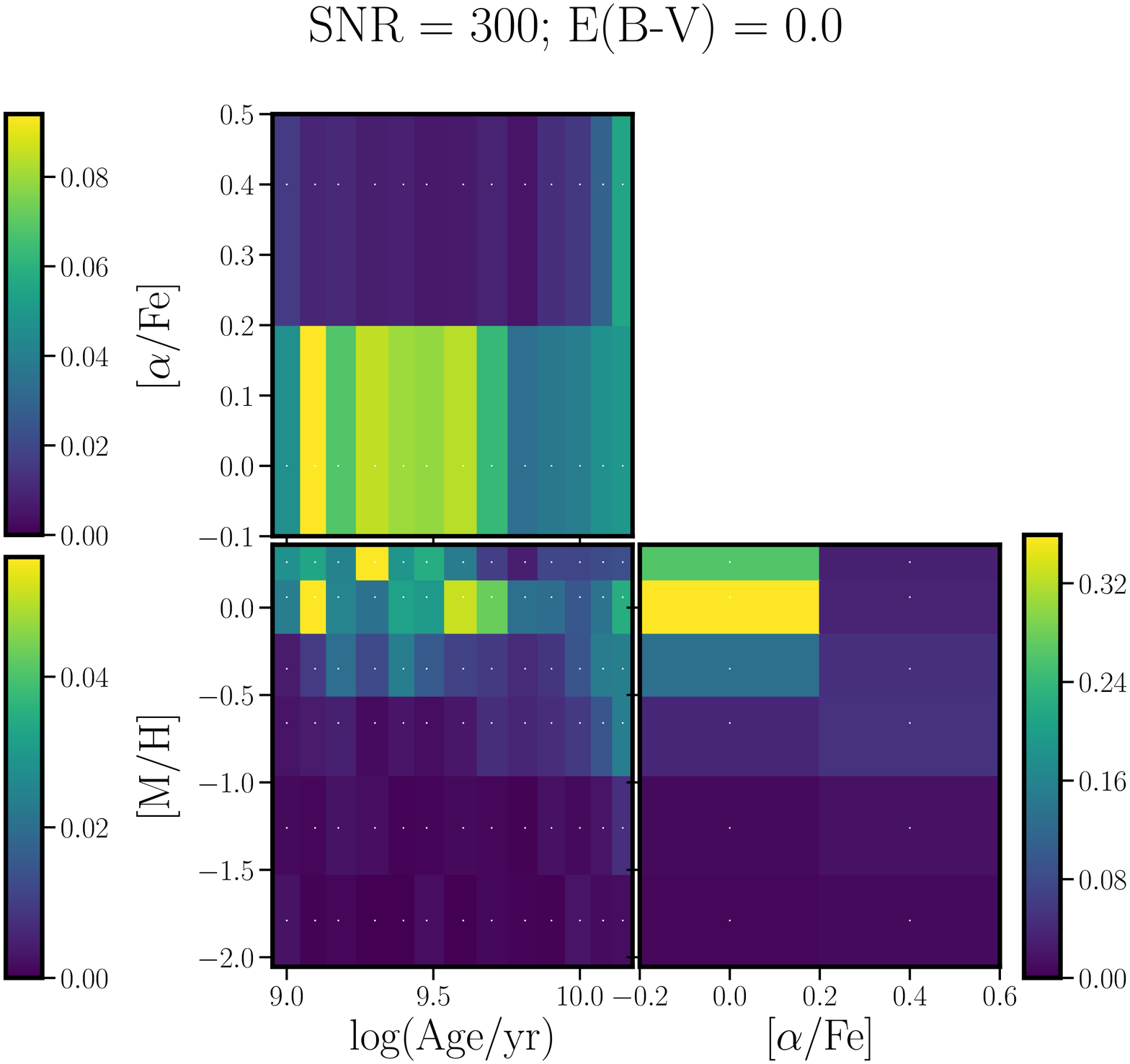}
 
 \caption{
 The left column displays the projected weightings of the templates that construct the mock spectra. The four rows correspond to the four compositions introduced in the text. The later three columns are the weighting maps we inferred from our fits of the mock spectra at S/N = 5, 60, and 300. The output maps are close to the inputs when S/N reaches 300, but still have difference.}
 \label{fig:mockW}
\end{figure*}

In order to test how the \ppxf\ fits work on different stellar population compositions at a range of S/N, we test our methods on mock spectra first. 
We make the mock spectra with four stellar population compositions by weighting the V15 model templates which are constructed on the BaSTI scaled-solar or \alf-enhanced isochrones with a Salpeter IMF. The left column in Figure~\ref{fig:mockW} displays the projected weighting maps of these four stellar population compositions. 

Two of them have one major component in each, mimicking single burst star formation histories. 
Both these two single major components are centred at log(age/Gyr) = log(6) and \ZH\ = 0.06 (the solar metallicity in V15 models). Their \afe\ is centred at 0 and 0.4 respectively. 
Each component has an 1\sgm\ Gaussian dispersion of ($\delta$(log(age/Gyr)), $\delta$(\ZH), $\delta$(\afe)) = (0.2, 0.3, 0.2). 

Each of the other two compositions contains two major components. 
One composition mimics ETGs, which contains a compact ``bulb" at a relatively old age, rich \ZH, and high \afe\ in the parameter space, plus a more extended younger component with lower \ZH\ and \afe. These two components are centred at (log(age/Gyr), \ZH, \afe) = (log(10), 0.2, 0.3) and (log(3.5), -0.5, 0.1), with the 1\sgm\ dispersion of ($\delta$(log(age/Gyr)), $\delta$(\ZH), $\delta$(\afe)) = (0.1, 0.15, 0.1) and (0.15, 0.3, 0.1), respectively. 

The other composition is more similar to late-type galaxies. It contains a bulge component with a rapid metal enrichment history plus a disk component with a mild and extended star formation history. The mean (log(age/Gyr), \ZH, \afe) of the bulge and disk components are (log(10), -0.66, 0.25) and (log(2), 0.06, 0). The 1\sgm\ dispersions ($\delta$(log(age/Gyr)), $\delta$(\ZH), $\delta$(\afe)) are (0.1, 0.5, 0.1) and (0.5, 0.3, 0.15) for the two components, respectively. 

\begin{figure}
 \includegraphics[width=\columnwidth]{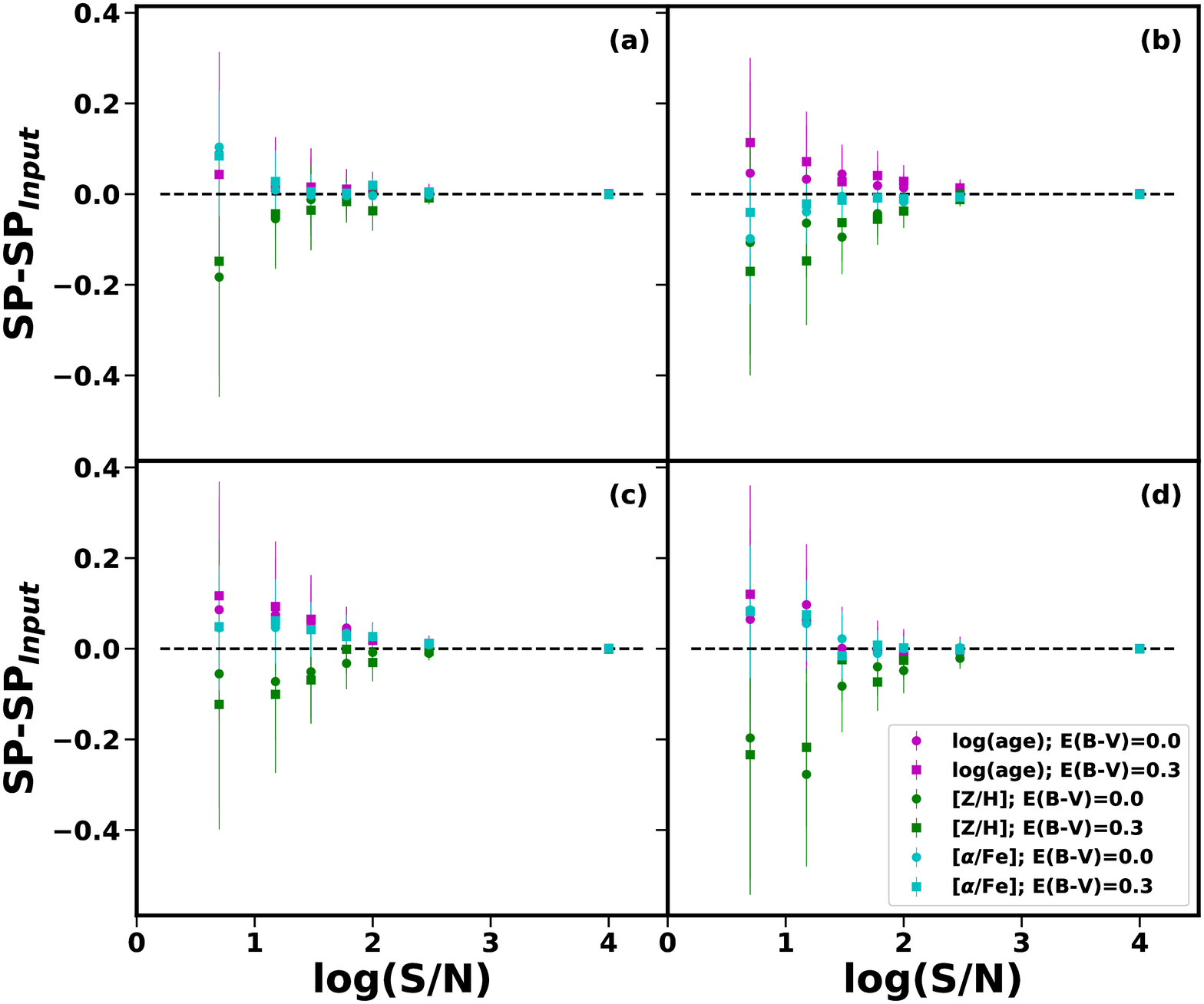} 
 \caption{
 The difference between the mean stellar population parameters from the fits of the mock spectra at different S/N and the parameters of the corresponding mock spectra, as functions of log(S/N). From (a) to (d), the four panels correspond to the four compositions displayed in Figure~\ref{fig:mockW}. The magenta, green, and cyan symbols represent log(age), [Z/H], and \afe, respectively. The circles and squares indicate the results from the spectra with E(B-V) = 0 and 0.3. 
In all cases, the differences converge to zero at S/N=10000. The differences at S/N$\sim$60 are acceptable for our study.}
 \label{fig:mockSNR}
\end{figure}

For each composition, we add constant noise and \citet{Calzetti_00} reddening curve to the combined mock spectra. The mean S/N we add are 5, 15, 30, 60, 100, 300, and 1000. The E(B-V) of our test spectra are either 0 or 0.3. 
Because the templates cannot be shrunk during the fitting procedures, we broaden our mock spectra to a velocity dispersion of 150~km/s. 

The fitting method is the same as the second fit in each iteration for our MaNGA spectra. The only difference is that the templates used in the fits are the original model templates without interpolations, which are the same as the ones that make the mock spectra. Thus the difference between the parameters of the mock spectra and our fitting results will be purely driven by the fitting methods and the noise of the spectra. 
For each mock spectrum, after the simultaneous fits for both its stellar populations and kinematics, we fix the kinematic results as inputs and preform 500 bootstrapping realisations to estimate the scatter of the stellar population parameters. 

We show the representative results of the weighting maps for the spectra at S/N = 5, 60, and 300 with E(B-V)=0 in Figure~\ref{fig:mockW}. Each row corresponds to one composition. From left to right, the first column displays the weighting maps of the mock spectra. The next three columns display the outputs at three S/N respectively. For all the four compositions, the results from S/N = 5 spectra do not show reasonable indications of age and \ZH. At S/N = 60, the fitting programmes may produce similar stellar population parameters to the input values, but fail in reconstructing the weighting maps. The input and output weighting maps become similar at S/N = 300, though not exactly the same. 
From the S/N distribution of our spectra (Figure~\ref{fig:histSNR}), our fitting procedures will not be able to tell the accurate stellar components. Therefore we do not show the weighting maps from our scientific analysis in this paper. 

Figure~\ref{fig:mockSNR} shows the difference between the input and output of the stellar population parameters as functions of log(S/N). From (a) to (d), the magenta, green, and cyan colours represent log(age), \ZH, and \afe, respectively. The circles and squares indicate the results from the spectra with E(B-V) = 0 and 0.3. 
In all cases, the differences converge to zero at S/N=10000. The differences at S/N$\sim$60 are below 0.05, which is acceptable for our study. 

\subsection{MaNGA Spectra}
\label{sec:Test:v10} 

\begin{figure}
 \includegraphics[width=\columnwidth]{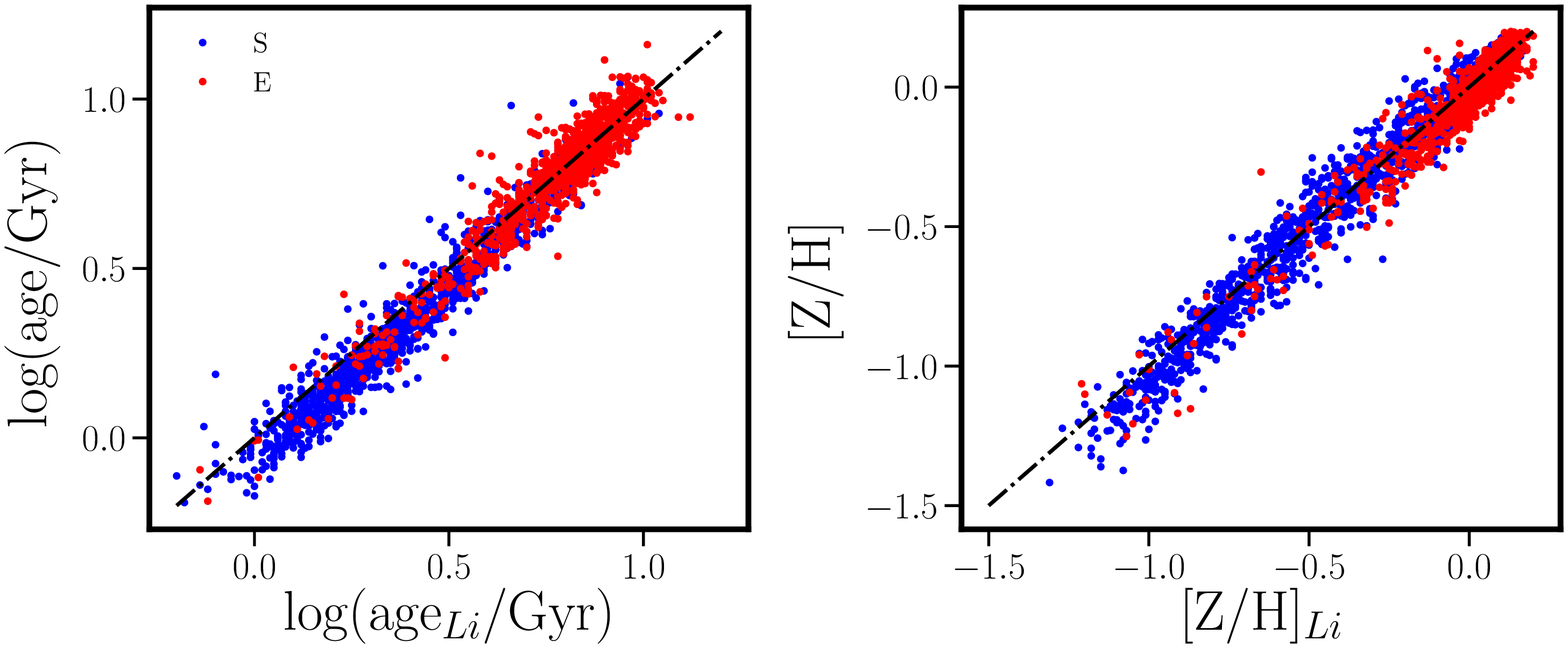}
 \caption{
 We measure the age and [Z/H] of the spectra stacked from 1\Rmaj\ using the same model templates and parameter settings as in L18. These plots show the comparison of our age (left) and [Z/H] (right) measurements with L18's. The red and blue points indicate the ETGs and late-type galaxies, respectively. The results from L18 and our work are in agreement.}
 \label{fig:compLiV10}
\end{figure}

L18 applied \ppxf\ to fit the ages and \ZH\ of MPL5 galaxies and provided the mean values within 1\re. They used the similar methods but only searched the best fits from the two dimensional parameter space of age and \ZH\ with the use of \citet[][hereafter V10]{V10} templates. 
To check if our programmes work correctly, we apply our programmes on the entire MPL5 galaxies but replacing the V15 templates with V10. Then we compare our results with L18's in Figure~\ref{fig:compLiV10}. Our parameter settings are the same as those in L18. 

In Figure~\ref{fig:compLiV10}, the x and y-axis indicate the the age (left) and \ZH\ (right) measurements from and L18 and our work, respectively. Our results are in good agreement, indicating that our spectral fitting programmes are reliable. 
Note that L18 measured the age and \ZH\ of the Voronoi bins within 1\re\ then averaged them respectively, while we estimate the parameters from the total light within 1\re. However, the results should be basically the same with such small difference. 

\subsection{Lick Index Analysis}
\label{sec:Test:LickI} 

\citet{Zheng_19} studied the \afe-\sgm\ relation based on the entire MPL7 sample of 2997 galaxies. They did the analysis with the use of TMJ11 model and four indices of H$_\beta$, Mg$b$, Fe5270, and Fe5335. Although they included both star-forming and passive galaxies, the passive galaxies (H$_\beta$ < 3) are significantly dominated in their massive sample and the star-forming and passive galaxies follow similar distributions of \afe\ and \sgm\ at low masses. Therefore we expect the \afe-\sgm\ relation shown in their work is similar to the relation fitted by the ETGs in their sample only. 

\begin{figure}
	\includegraphics[width=\columnwidth]{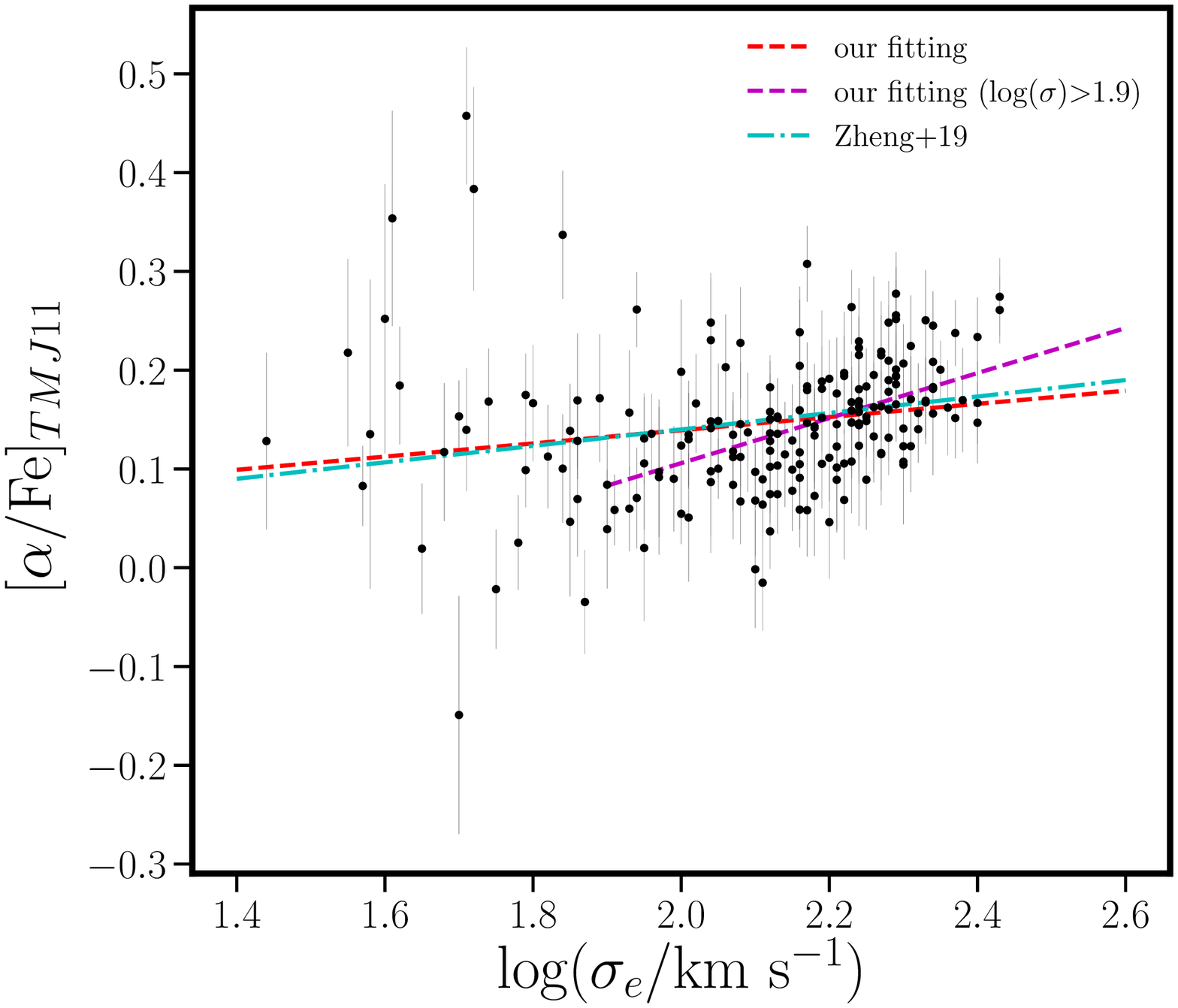}
    \caption{
    The \afe\ of our sample ETGs measured in the same way as used in \citet{Zheng_19}. It is fitted by four Lick indices H$_\beta$, Mg$b$, Fe5270, and Fe5335, with the use of TMJ11 model. 
    The red dashed line and the cyan dot-dashed line are the \afe-\sgm\ relations fitted from our results and \citet{Zheng_19}, respectively. The slopes of the two relations are 0.06 and 0.08, basically the same. The magenta dashed line is the relation fitted from our results of the massive ETGs, with a positive slope of 0.22.}
    \label{fig:TMJ11}
\end{figure}

To check the representativeness of our high-S/N sample, we repeat the analysis in \citet{Zheng_19} on the spectra stacked from the inner 1\Rmaj\ ellipses of our sample ETGs and compare our resulted \afe-\sgm\ relation with theirs. 
In Figure~\ref{fig:TMJ11} we plot our \afe\ measurements against log(\sgmReKms). The red and magenta dashed lines are the \afe-log(\sgm) relations fitted from our results of the entire and massive samples respectively. The cyan dot-dashed line is the \afe-log(\sgm) relation adopted from the Figure~17 in \citet{Zheng_19}. The slopes of the red dashed line and the cyan dot-dashed line are 0.06 and 0.08, respectively. The small difference indicates that our sample is well representative. 
When limiting to the massive sample (log(\sgmReKms) > 1.9), the slope of the \afe-log(\sgm) relation is 0.22, agree with the results from the literature. 

\begin{figure}
	\includegraphics[width=\columnwidth]{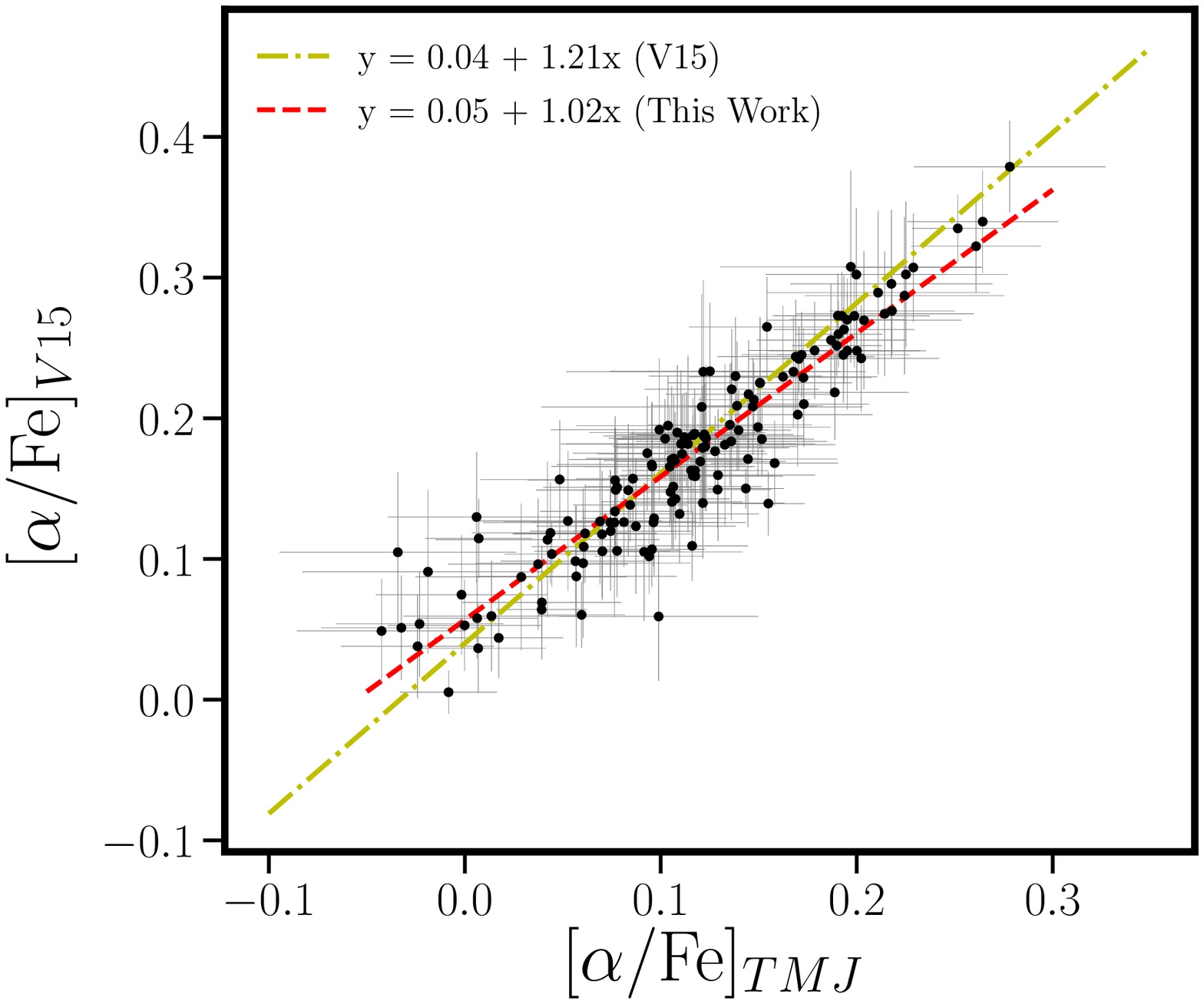}
    \caption{
    The comparison of the \afe\ measurements that are derived from different stellar population models. The data points are our sample ETGs. The x- and y-axis indicate the \afe\ which is fitted by V15 and TMJ11 models respectively, with the use of the Lick index analysis that is described in V15 paper. 
    The red dashed line is the relation between the \afe\ derived from the two approaches based on our sample. V15 provided the relation they fitted by the measurements of 4671 nearby ETGs, which is shown by the yellow dot-dashed line. The two relations are basically the same, indicating that our relatively small sample is representative. 
     }
    \label{fig:TMJ_V3I}
\end{figure}

V15 showed another test with Lick index analysis. They estimated the ``true \afe" of 4671 nearby ETGs by fitting their \afe\ and \ZH\ from three indices \mgb, 
[MgFe]'=$\sqrt{Mgb\times(0.72Fe5270+0.28Fe5335)}$, and Fe3=(Fe4383+Fe5270+Fe5335)/3 at fixed ages. The fixed ages were estimated by the spectral fitting method with V10 templates. Their results from the fits with V15 and TMJ11 models followed a relation as \afe$_{V15}$=0.04+1.21\afe$_{TMJ11}$. 
We apply the same analysis on the spectra stacked from the inner 1\Rmaj\ of our sample ETGs. Figure~\ref{fig:TMJ_V3I} shows the results. The relation that are fitted from our results is \afe$_{V15}$=0.05+1.02\afe$_{TMJ11}$ (red dashed line), consistent with the relation that V15 derived (yellow dot-dashed line) within uncertainties.


\section{Results}
\label{sec:Result}

\subsection{The Mean \afe\ within 1\re}
\label{sec:1Re}

\begin{figure*}
	\includegraphics[width=0.9\textwidth]{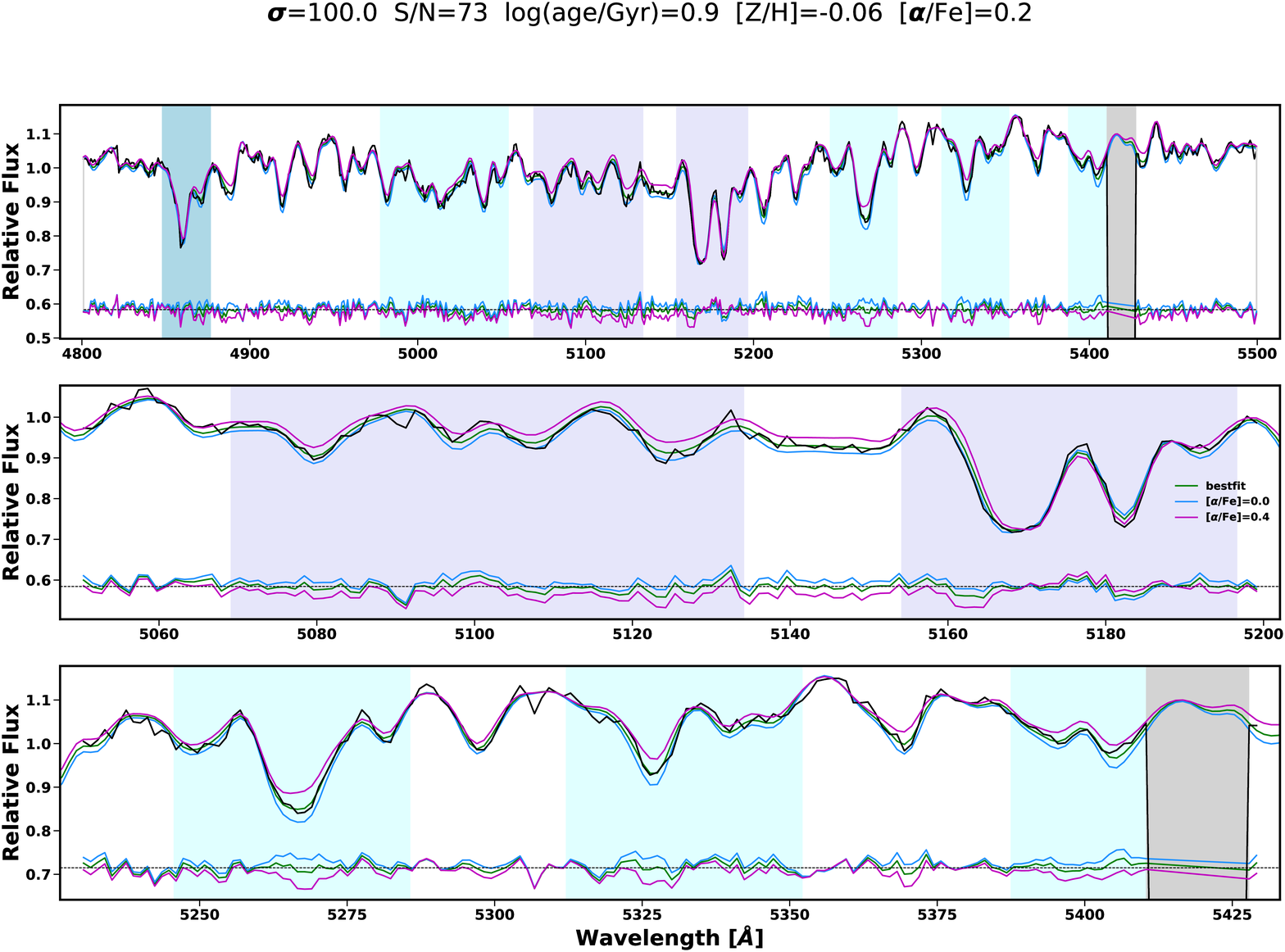}
	\includegraphics[width=0.9\textwidth]{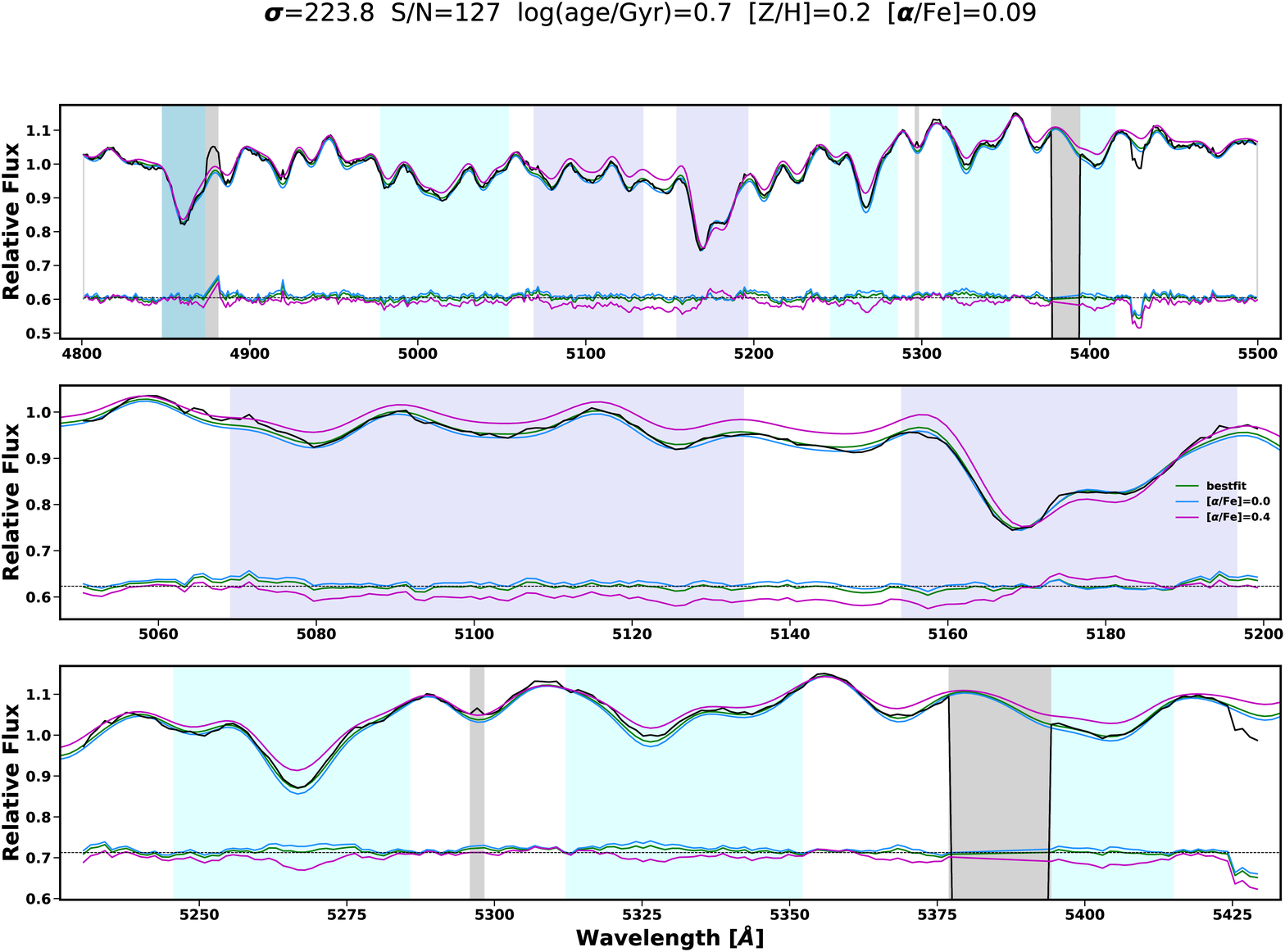}
    \caption{
    The fits of spectra stacked from the inner 1\Rmaj\ ellipses of two representative ETGs with different \sgmRe. The related parameters are labeled on the top of each panel (see text for the detail). 
    In each panel, the top row displays the fits of the spectrum over  4800-5500\AA. The middle and bottom rows are the zoom-in plots for Mg and Fe bands respectively. The black and green lines are the galactic spectra and the best fitting results. The best fits match the spectra precisely for both ETGs. Other details are provided in the text. 
     }
    \label{fig:fitSpec}
\end{figure*}

We stack the spectrum within the inner 1\Rmaj\ ellipse of each ETG in our sample and fit its \afe\ by \ppxf. 
In Figure~\ref{fig:fitSpec} we show the spectral fits of two representative ETGs with different \sgmRe. 
On the top of each panel we label the related parameters, among which the S/N and three stellar population parameters are derived from the fits. 
In each panel, the top row displays the fits of the spectra over the entire wavelength range of 4800-5500\AA. The middle and bottom rows are the zoom-in plots for Mg and Fe bands respectively. 
The black and green lines are the galactic spectra and the best fitting results. The cyan and magenta spectra are constructed by the model templates with the same weights in age and \ZH\ dimensions as the best fit's. However, their \afe\ are forced to be 0 or 0.4 respectively. 
The lower flat lines with fluctuations are the residuals from the fit. The colours correspond to the three colours of the fitting spectra. The grey horizontal line is the baseline for the residuals. 

The blue bands illustrate the index bandpass of H$_\beta$. The magenta and cyan bands mark the \alf-element and Fe indicators respectively. The three indices that indicate the \alf-elements in this wavelength range are \mgI, Mg$_2$, and \mgb. The Mg$_2$ and \mgb\ indices are overlapped with each other. The four Fe tracers are Fe5015, Fe5270, Fe5335, and Fe5406. The grey bands are the regions which are excluded from our fits. 

For both ETGs with different \sgmRe, the best fits match the spectra precisely. The cyan and magenta model spectra with different \afe\ fail in fitting the Mg and Fe features. The magenta spectra with a higher \afe\ have stronger \mgb\ and weaker Fe absorption features. The cyan spectra show the offset in the opposite direction. These indicate that our spectral fitting programmes work well. Especially, they are able to tell the difference in \afe. 

Particularly, the magenta and cyan spectra are above and below the green spectra at \mgI\ features respectively. It indicates that the V15 templates with higher \afe\ have lower \mgI\ index values and vice versa. V15 mentioned such unusual behaviour of \mgI. We have further discussions in Section~\ref{sec:mg1}. 

\begin{figure}
	\includegraphics[width=\columnwidth]{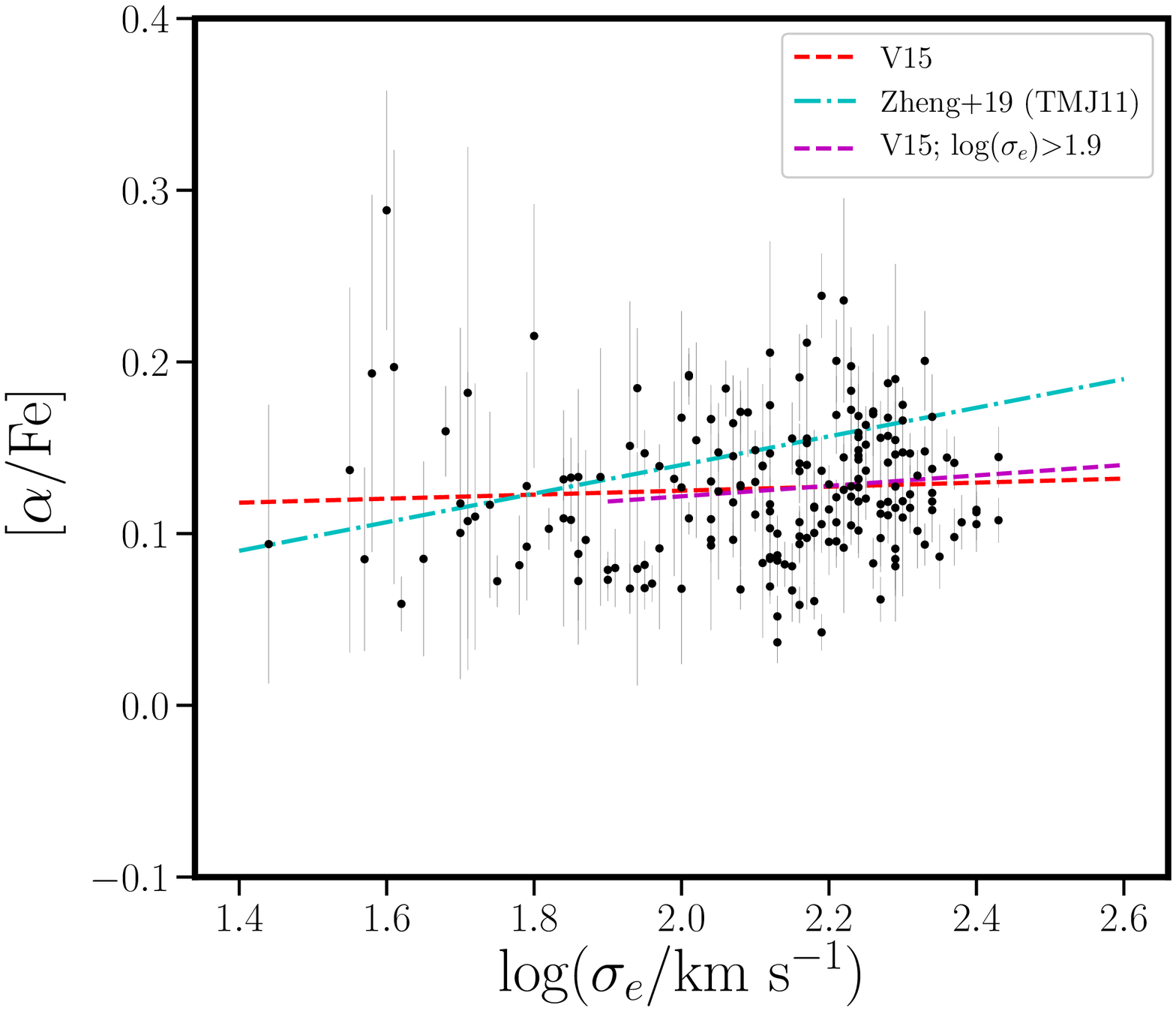}
    \caption{
    The relation between the \afe\ from our fits and log(\sgmReKms) of our sample ETGs. The red and magenta dashed lines are the linear fits for our entire and massive (log(\sgmReKms) > 1.9) samples respectively. Their slopes are 0.01 and 0.03. The cyan dot-dashed line is the \afe-\sgm\ relation from \citet{Zheng_19} (the cyan dot-dashed line in Figure~\ref{fig:TMJ11}). 
    The relations based on our \afe\ measurements are flat and do not support the classic conclusion of a positive \afe-\sgm\ relation of ETGs. 
     }
    \label{fig:afe_sgm}
\end{figure}

Figure~\ref{fig:afe_sgm} presents our fitting results of \afe\ as a function of log(\sgmReKms). The error bars show the 1\sgm\ scatter from the bootstrapping fits with a 500 times realisation. 
Similar to the \afe-\sgm\ relations shown in \citet{Liu_16c} and \citet{Zheng_19}, the scatter of \afe\ is relatively large at low masses. Unfortunately, we do not have enough low-mass ETGs for statistical analysis. 

The red and magenta dashed lines are the fitted relations for our entire and massive (log(\sgmRe/km s$^{-1}$) > 1.9) samples respectively. For reference, we plot the \afe-log(\sgm) relation from \citet{Zheng_19} as the cyan dot-dashed line (the same as the cyan dot-dashed line in Figure~\ref{fig:TMJ11}). 
The slopes of the red and magenta dashed lines are 0.01 and 0.03, respectively. The flatness of these relations indicates that the \afe\ of ETGs from our analysis does not have significant dependence on galactic potential wells, even if limiting to the massive ETGs only. 

At the same time, however, the results from our analysis with TMJ11 model support the classic positive \afe-\sgm\ relation of ETGs. From Section~\ref{sec:Test:LickI}, the slopes of the TMJ11-based \afe-log(\sgmRe) relations of the same entire and massive samples (the red and magenta dashed lines in Figure~\ref{fig:TMJ11}) are 0.06 and 0.22 respectively. 
We will further discuss it in Section~\ref{sec:Discussion}. 

\begin{figure}
	\includegraphics[width=\columnwidth]{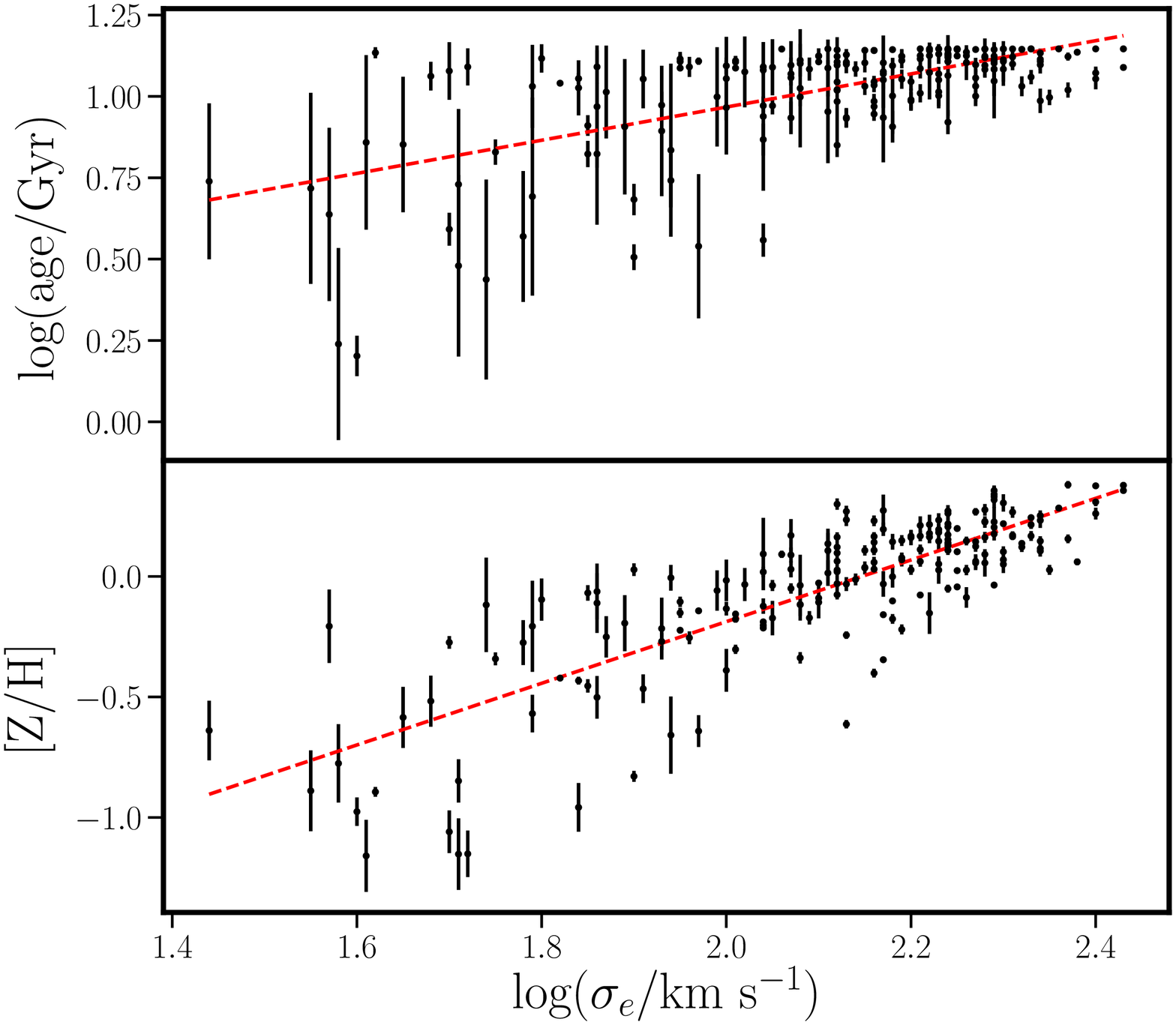}
    \caption{
    The upper and lower panels present the age and \ZH\ measurements from our fits as functions of log(\sgmReKms), respectively. The red dashed lines illustrate the linearly fitted relations. The slopes are 0.51 and 1.28 in the upper and lower panels. However, the former number is unreliable because the age estimations of a significant fraction of our sample ETGs are limited by the model's upper boundary. 
     }
    \label{fig:ageZ_sgm}
\end{figure}
\begin{figure}
	\includegraphics[width=\columnwidth]{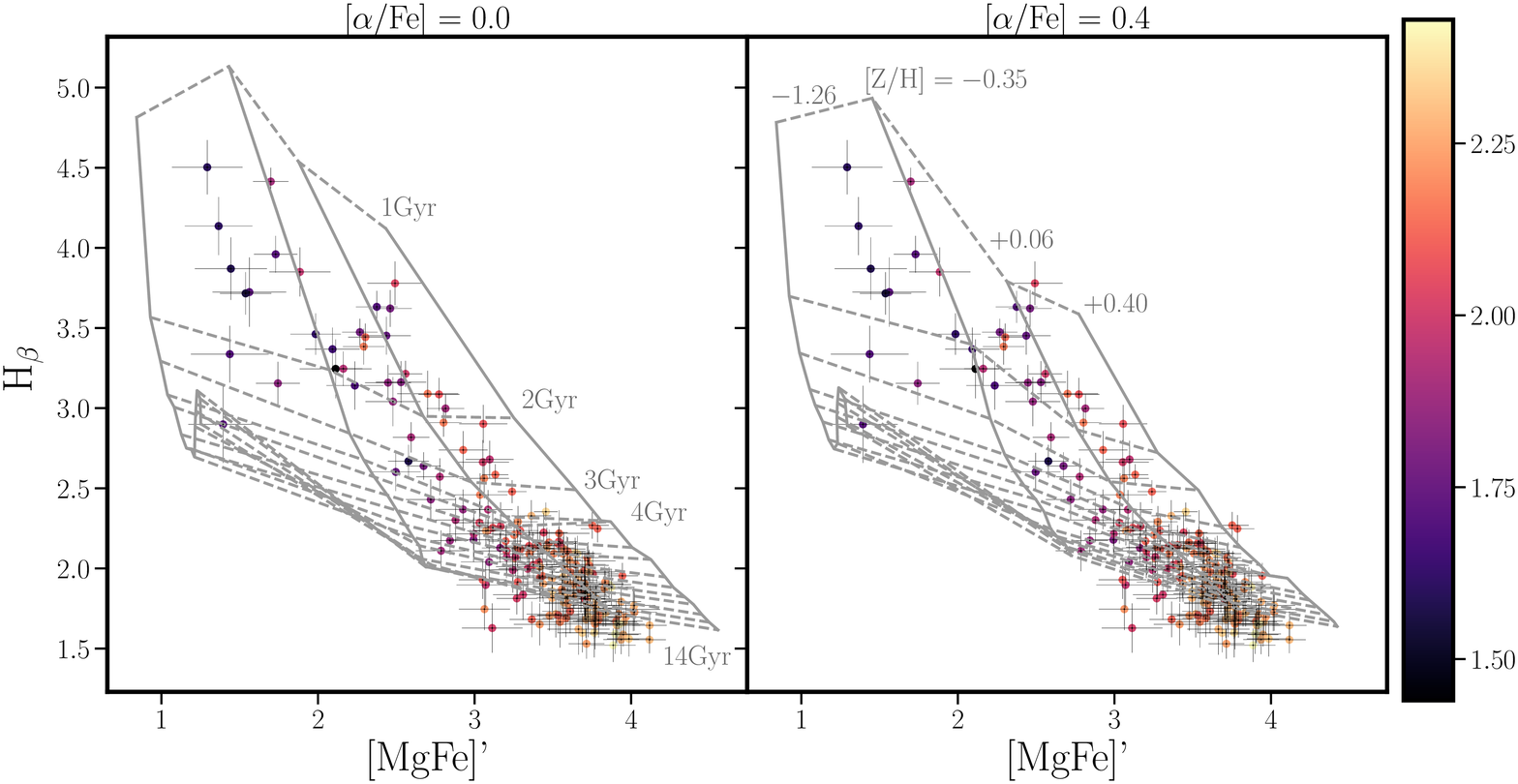}
    \caption{
    The V15 model grids at \afe\ $=$ 0 (left) and 0.4 (right) in H$_\beta$-[MgFe]' planes. The dashed lines are the isolines of the ages from 1 to 14~Gyr, with an interval of 1~Gyr. The solid lines illustrate the isolines of \ZH\ $=$ $-$1.26, $-$0.35, 0.06, and 0.4. The data points of our sample ETGs are colour coded according to their log(\sgmReKms). The colour bar is displayed on the right. A significant fraction of our sample ETGs have their H$_\beta$ lower than the model boundary.
     }
    \label{fig:HbMgFe_grid}
\end{figure}

For reference, we show our measurements of age and \ZH\ as functions of log(\sgmReKms) in Figure~\ref{fig:ageZ_sgm}. Both the log(age) and \ZH\ have positive correlations with log(\sgmRe), in agreement with the results from the literature. The slopes of our log(age)-log(\sgmRe) and \ZH-log(\sgmRe) relations are 0.51 and 1.28. However, the former number is unreliable, for a significant fraction of our sample have their age estimations limited by the model's upper boundary. 

In Figure~\ref{fig:HbMgFe_grid}, we plot the V15 model grids in H$_\beta$-[MgFe]' planes. The data points of our sample ETGs are colour coded according to their log(\sgmReKms). A significant fraction of our sample have their H$_\beta$ lower than the model boundary, indicating that the oldest age in V15 models, 14~Gyr, is not enough for our oldest ETGs. As a result, the age measurements of these ETGs are limited by the model's upper boundary, which are untrue.

\subsection{Radial Gradients within 1\re}
\label{sec:grad}

\begin{figure}
	\includegraphics[width=\columnwidth]{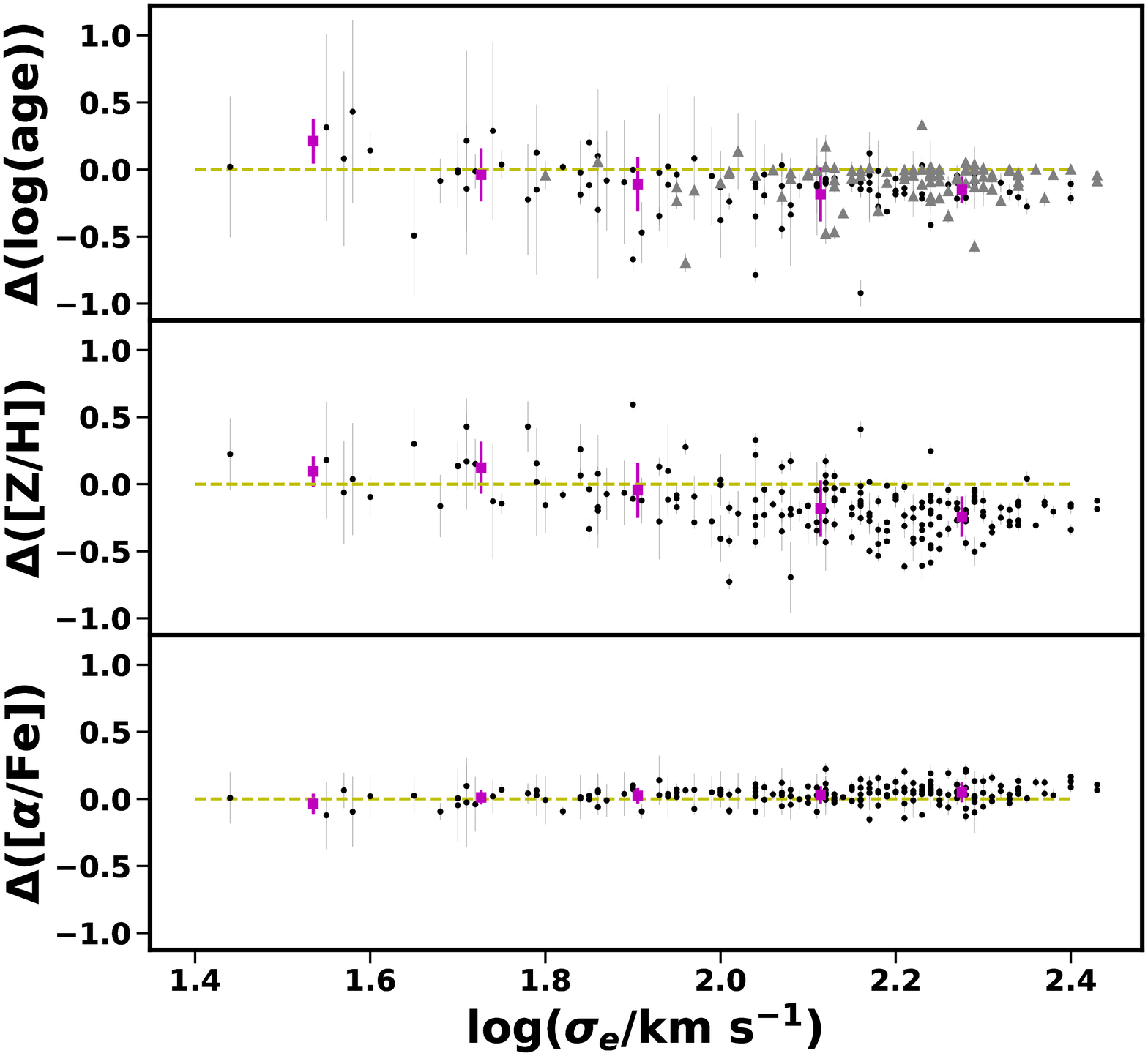}
    \caption{
    The top, middle, and bottom panels present the radial gradients of log(age), \ZH, and \afe\ of our sample ETGs as functions of log(\sgmReKms), respectively. The grey data points in the top panel indicate the ETGs which have ages reaching the model boundary in more than one radial bins. 
    The yellow dashed lines mark the zero gradients. The magenta points and error bars show the mean gradients and 1\sgm\ scatter in different bins of log(\sgmRe). The grey data points are not taken into account in the mean calculations. 
    On average, the age and \ZH\ gradients have similar trends with log(\sgmRe), but age profiles are much flatter at each \sgmRe. The \ZH\ gradients are positive at low \sgmRe\ and gradually decrease to negative in the high-mass range. They are more negative at higher \sgmRe. The \afe\ gradients are flat over the full range of log(\sgmRe), but slightly turn to positive at the massive end. 
     }
    \label{fig:Grad_sgm}
\end{figure}
\begin{figure*}
	\includegraphics[width=\textwidth]{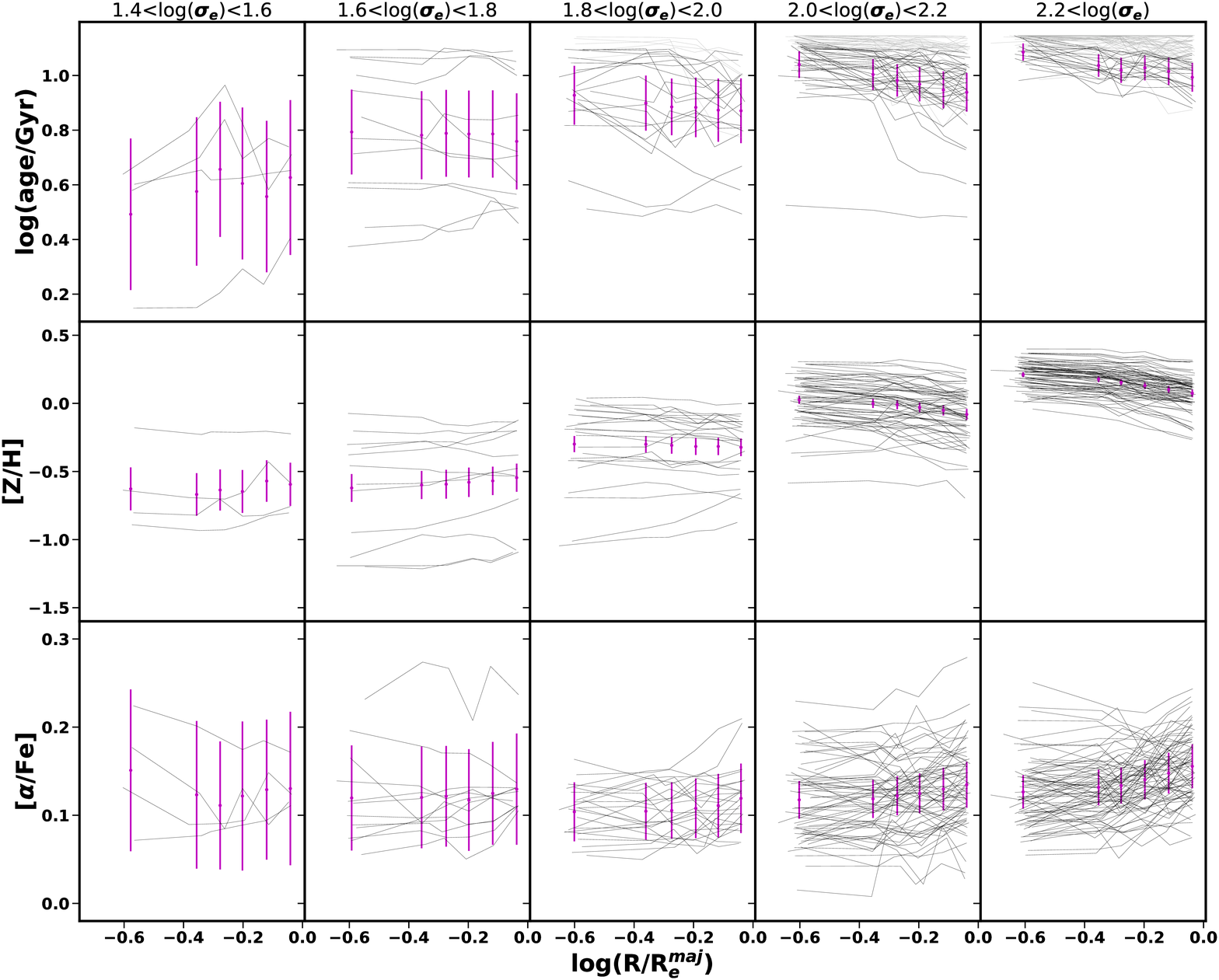}
    \caption{
    The radial profiles of log(age/Gyr) (the top row), \ZH\ (the middle row), and \afe\ (the bottom row) of the individual ETGs in the log(\sgmReKms) bins that are applied in Figure~\ref{fig:Grad_sgm}. The grey profiles in the top panels correspond to the grey data points in Figure~\ref{fig:Grad_sgm}. The magenta dots and error bars illustrate the mean radii, mean log(age/Gyr), \ZH, or \afe\ and the mean 1\sgm\ uncertainties of them within the radial bins. The grey profiles are not taken into account in the mean calculations.
     }
    \label{fig:profBin}
\end{figure*}

Figure~\ref{fig:Grad_sgm} presents the radial gradients of log(age), \ZH, and \afe\ within 1\Rmaj\ of our sample ETGs. The error bars show the 1\sgm\ scatter from the Monte Carlo simulation with a 500 times realisation. The magenta symbols show the mean gradients and 1\sgm\ scatter in every 0.2 interval of log(\sgmReKms) starting from 1.4. Their error bars show the scatter of the gradients within the log(\sgmRe) bins. 
Figure~\ref{fig:profBin} presents the log(age/Gyr), \ZH, and \afe\ profiles of the individual ETGs in different log(\sgmRe) bins. The magenta dots and error bars illustrate the mean radii, mean log(age/Gyr), \ZH, or \afe, and the mean 1\sgm\ uncertainties of them within the radial bins. In both figures, the grey symbols in the top panel indicate the ETGs which have ages reaching the model boundary in more than one radial bins, and they are not taken into account in the mean calculation. 

On average, the age and \ZH\ gradients have similar trends with log(\sgmRe), but the age profiles are much flatter at each \sgmRe. The \ZH\ gradients are positive at low \sgmRe. They gradually decrease to negative in the high-mass range. 
The \afe\ gradients are flat when log(\sgmReKms) is below 2.1. This agrees with the previous results, e.g., \citet{Mehlert_03,Kuntschner_04,Greene_15,Alton_18,Parikh_19}. At higher \sgmRe, it slightly turns to positive. 
\citet{Goddard_17b} measured the age and metallicity gradients of the ETGs from the MaNGA MPL4 sample. They employed the \texttt{FIREFLY} spectral fitting code \citep{MarastonStromback_11} to analysis the stellar populations and divided the space by Voronoi binning method. Though their results of age gradients are generally zero, they are consistent with ours within error bars. Their metallicity gradients are similar to ours, which are, on average, positive at low masses, and gradually decrease to zero and then negative as the increase of galactic masses. 
Based on the CALIFA data, \citet{MartinNavarro_18} stacked 45 massive ETGs with \sgmRe\ > 150~km/s in three \sgmRe\ bins and made their stellar population radial profiles within 1\re. Their \ZH\ gradients are negative, and the age and \afe\ gradients are close to zero. These are all in agreement with our results. 

These results indicate that massive ETGs have higher metallicity, slightly older ages and lower \afe\ in galactic central regions. The differences between central and outer regions get larger as galactic mass increases. 
For the flat age gradients, probably because in galactic centres the star formation processes start earlier and last longer, and the balance of old and young stars makes the mean stellar ages in galactic centres similar to the outer regions. The slightly more old or young stars in central regions would make the gradients little negative or positive. 

Regarding to the \ZH\ and \afe\ gradients, in massive ETGs, the reasons might be that they have deeper gravitational potential wells in the centre therefore have higher gas content to support more extended star formation activities than in outskirts. 
In intermediate-mass galaxies, though the central regions still have more intense star formation activities than outside, the metal enriched gas in the centres are easier to be blown out and fall into outer regions comparing to the most massive galaxies. Thus less massive ETGs have flatter profiles. 

In the low-mass range, while the \afe\ profiles are flat, a significant fraction of galaxies have positive metallicity gradients, indicating a lower metallicity in galactic centres. A possible scenario is that the star formation activities of low-mass galaxies do not vary too much from centre to outside. At the same time, galactic central regions have more infall gas with relatively low metallicity. When the central metal enrichment cannot overcome the metallicity dilution caused by gas infall, there will be a positive metallicity gradient. 

A caveat is that the uncertainty of these radial gradient estimations cannot be neglected, especially for ages and \afe. From Figure~\ref{fig:HbMgFe_grid}, the resolution of age estimations is low when it is older than 7~Gyr. Thus the age radial profiles could significantly change after considering the uncertainties at old ages. 
In Section~\ref{sec:mg1}, we demonstrate that V15 models might underestimate the \afe\ of some massive ETGs. Such underestimation might flatten the intrinsically positive \afe-\sgm\ relation and make the \afe\ distribution flat in Figure~\ref{fig:afe_sgm}. At the same time, it could also flatten the \afe\ radial profiles.


\section{Discussion}
\label{sec:Discussion}

In Section~\ref{sec:1Re}, Figure~\ref{fig:afe_sgm} shows that the \afe\ of ETGs, even if only limiting to the massive objects, does not correlate with \sgmRe\ when we estimate the \afe\ by spectral fitting method with V15 templates. It conflicts with the classic result of a positive \afe-\sgm\ relation of massive ETGs from the literature. 
However, based on the same sample of ETGs, there is a positive \afe-\sgm\ relation when we estimate the \afe\ by Lick index analysis with TMJ11 model (Section~\ref{sec:Test:LickI}, Figure~\ref{fig:TMJ11}). 
In this section, we further discuss the reasons for the different \afe\ measurements from different approaches. 

\subsection{Weighting Templets vs. Single-equivalent Stellar Population}
\label{sec:compMethods}

Using \ppxf, our spectral fitting method looks for the best weightings of the templates with different stellar population parameters. Essentially, it looks for the best combinations of different stellar population components. The results are the composite stellar population (CSP) parameters. 
However, our index analysis works in a different way. It searches the best matched single-equivalent stellar population (SSP) parameters from the model grids. 

To investigate the difference between CSP and SSP results, we interpolate the V15 templates according to the ages, \ZH, and \afe\ in the interpolated V15 model grid of Lick indices we introduce in Section~\ref{sec:3I}. Then we apply \ppxf\ to fit our galactic spectra that are stacked from the inner 1\Rmaj\ ellipses with a single interpolated template at one time. Thus we are searching the best fitting SSP templates. The parameter settings and the wavelength range for our fits are the same as what we adopt for the analysis in Section~\ref{sec:1Re}. 

\begin{figure*}
	\includegraphics[width=\textwidth]{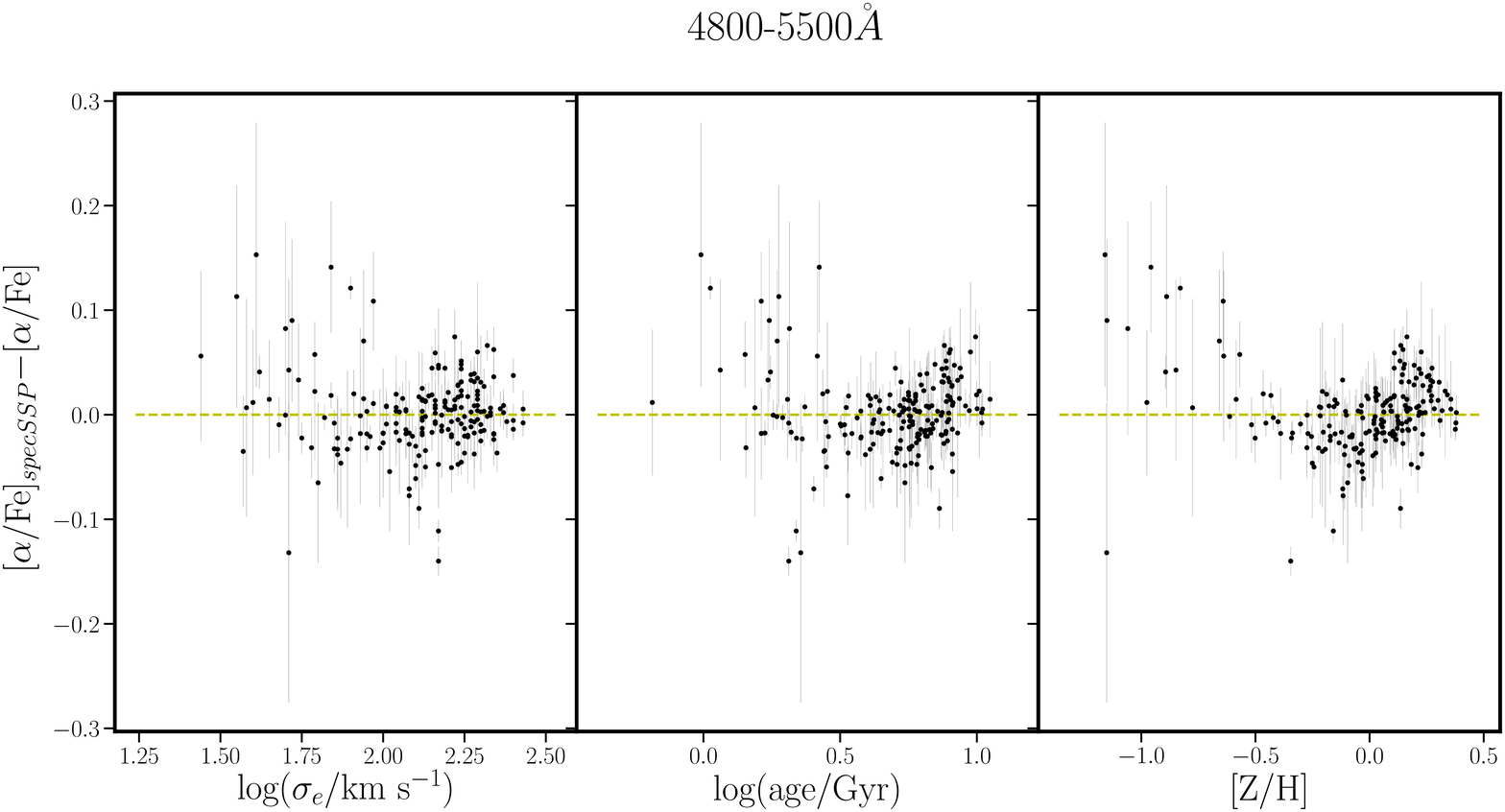}
    \caption{
    The y-axis indicates the difference of the mean \afe\ within 1\Rmaj\ ellipses of our sample ETGs between SSP and CSP spectral fitting results. From left to right, we show it as functions with log(\sgmReKms), log(age/Gyr), and \ZH. The yellow dashed lines show the zero difference. 
    The SSP results of \afe\ is significantly higher than the CSP results at low \sgmRe, young age, and low \ZH. The differences are correlated with \ZH\ best. Below \ZH$=-$0.5, they are larger at lower \ZH. Above \ZH$=-$0.4, the differences show a positive correlation with \ZH. 
     }
    \label{fig:dlt_SSPcsp}
\end{figure*}
\begin{figure*}
	\includegraphics[width=\textwidth]{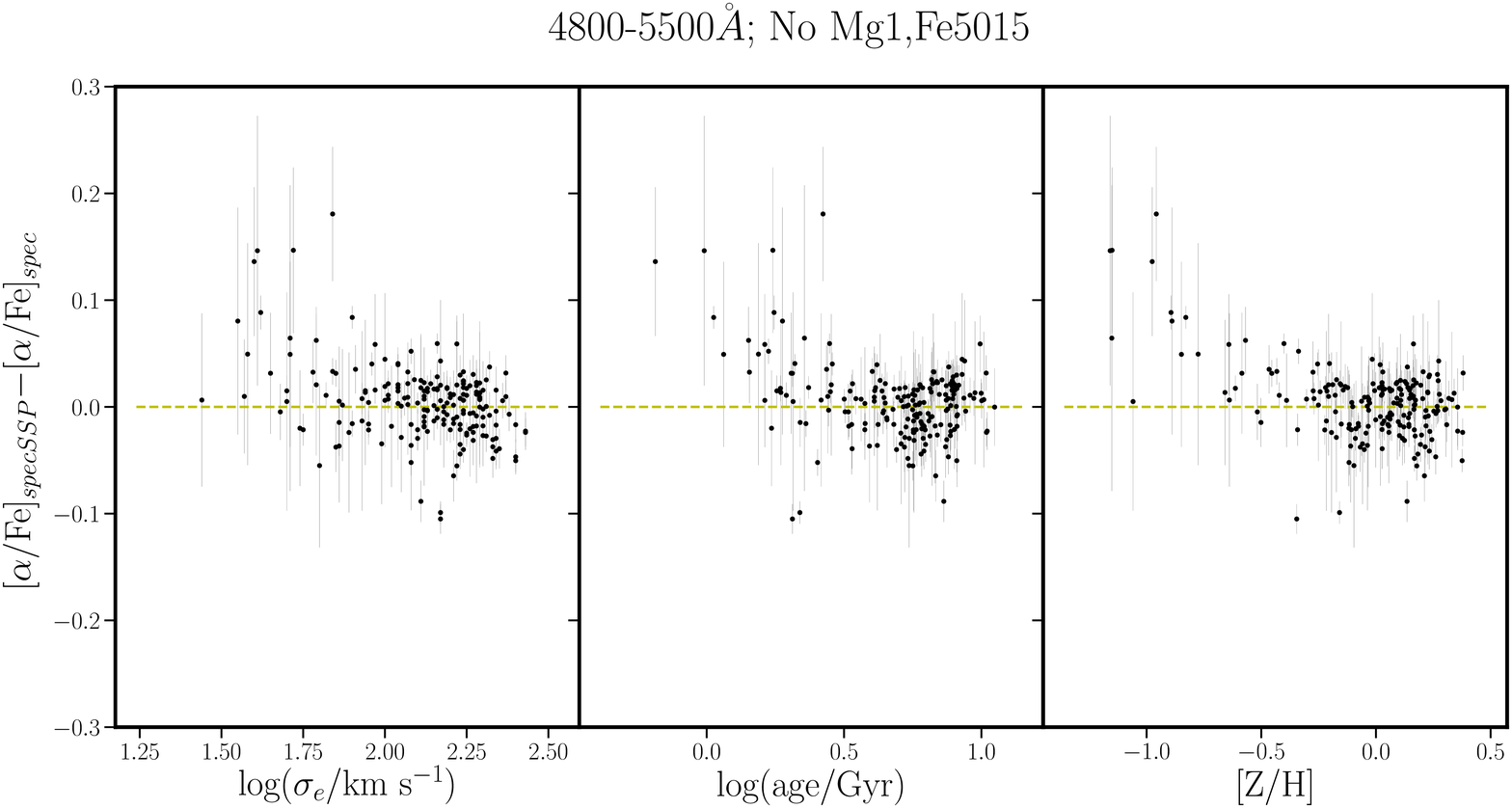}
    \caption{
    Same as Figure~\ref{fig:dlt_SSPcsp}, but the SSP and CSP fits for \afe\ estimations do not include the \mgI\ and Fe5015 bands. Comparing to Figure~\ref{fig:dlt_SSPcsp}, the differences between SSP and CSP fitting results of \afe\ do not correlate with log(\sgmRe) and \ZH\ in the high-mass range or at \ZH$>-$0.3. 
     }
    \label{fig:dlt_SSPcsp_MgFe}
\end{figure*}

Figure~\ref{fig:dlt_SSPcsp} presents how the \afe\ measurements change after we switch the spectral fits into SSP mode. The y-axis is the difference between the SSP results and our CSP results which Figure~\ref{fig:afe_sgm} shows. From left to right, the x-axis indicates log(\sgmReKms), log(age/Gyr), and \ZH. The \ZH\ is derived from the same spectral fitting analysis for our \afe\ estimations. 
In Section~\ref{sec:1Re} we mention that a significant fraction of our sample have their age estimations reach the model boundary of 14~Gyr. To avoid this problem, the age measurements are from the spectral fits with V10 templates. 
The yellow horizontal dashed lines mark the zero difference. 

In the high-mass range, the differences are relatively small (below 0.1~dex). At low \sgm, the SSP results are systematically higher. 
These changes do not change our result of a flat \afe-log(\sgmRe) relation of ETGs, especially at massive end. The slopes of the SSP based \afe-log(\sgmRe) relations for the entire and massive samples are 0.0 and 0.07. 

The differences correlate with \ZH\ best. 
Below \ZH$=-$0.5, the difference becomes larger at lower \ZH. It is probably due to the degeneracy between low \ZH\ and high \afe, and the CSP fits have an advantage in breaking the degeneracy. 
Above \ZH$=-$0.4, the \afe\ differences show a positive correlation with \ZH. However, this is because of the inclusion of the wavelength ranges for \mgI\ and Fe5015. 

According to the Figure~13 in V15, \mgI, which is traditionally regarded as an indicator of \alf\ elements, has unusual behaviours. When the ages are older than 1~Gyr, at a same \ZH, the V15 templates with higher \afe\ have lower \mgI\ index values. The reason is possibly that \mgI\ is more sensitive to C than Mg. Besides, because Fe5015 is sensitive to both Fe and Ti, an \alf\ element, it shows less sensitivity to the change of \alf\ element abundance in the models comparing to other Fe indicators. 

We mask the wavelength ranges of \mgI\ and Fe5015 and repeat both the SSP and CSP fits for our sample ETGs. Figure~\ref{fig:dlt_SSPcsp_MgFe} is similar to Figure~\ref{fig:dlt_SSPcsp} but the y-axis is the difference between the SSP-\afe\ and CSP-\afe\ that are measured under the masks. 
In this figure, the difference is still anti-correlated with \ZH\ at low \ZH, but the dependence with \ZH\ disappears above \ZH$=-$0.5.

\subsection{Full Spectra vs. Key Spectral Features}
\label{sec:compSpecI}

\begin{figure*}
	\includegraphics[width=\textwidth]{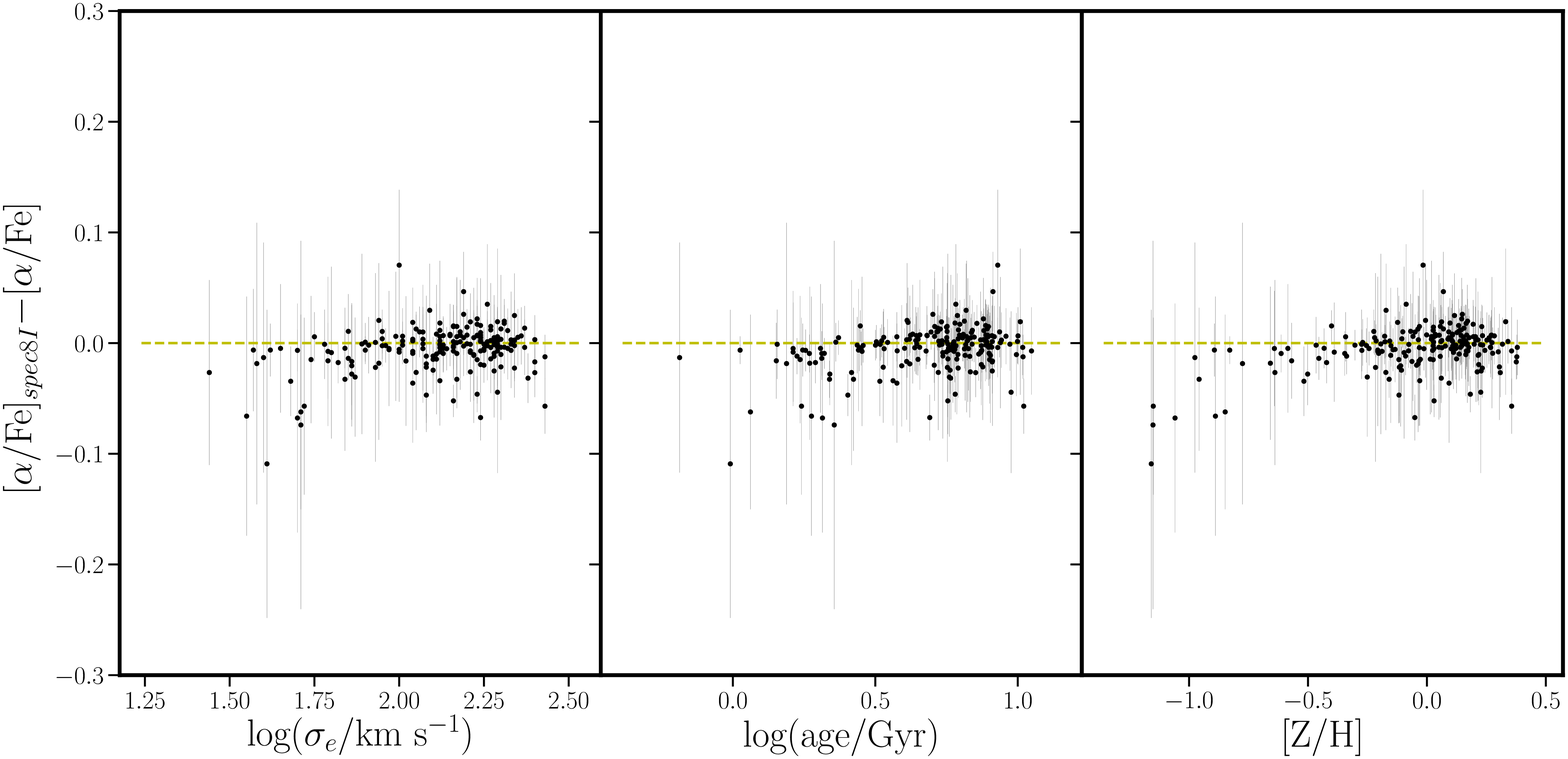}
    \caption{
    Similar to Figure~\ref{fig:dlt_SSPcsp}, but the y-axis is the difference between the spectral fits with different wavelength ranges. We fitted the spectra with the bands of the 8 Lick indices within 4800-5500\AA\ only. The y-axis is the difference between the resulted \afe\ and our main results. The \afe\ measured from two wavelength ranges are consistent with each other within the uncertainties, indicating that the non-prominent spectral features do not make significant difference in stellar population analysis. 
     }
    \label{fig:dltV_afeSpec8I}
\end{figure*}

Another key difference between our spectral fits and index analysis is that they take information from different spectral features. While the former takes into account the flux at every wavelength, the latter only makes use of the most prominent features within the wavelength range. 
In this section we examine how much difference the non-prominent features would make in the \afe\ measurement. 

The wavelength range for our spectral fits (4800-5500\AA) covers 8 Lick indices: H$_\beta$, Fe5015, \mgI, Mg$_2$, \mgb, Fe5270, Fe5335, and Fe5406. We apply the same spectral fitting methods as we do for the main scientific results on the spectra stacked from the inner 1\Rmaj\ ellipses of our sample ETGs, but limiting the wavelength range to the bands of these Lick indices only. Then we compare the resulted \afe\ to our main results.  
Similar to Figure~\ref{fig:dlt_SSPcsp}, Figure~\ref{fig:dltV_afeSpec8I} shows the difference as functions of log(\sgmReKms), log(age/Gyr), and \ZH. For all of our ETGs, the \afe\ measured from two wavelength ranges are consistent with each other within the uncertainties. It indicates that over 4800-5500\AA, the non-prominent spectral features do not make significant difference in stellar population analysis. 

\subsection{\mgI}
\label{sec:mg1}

\begin{figure*}
	\includegraphics[width=\textwidth]{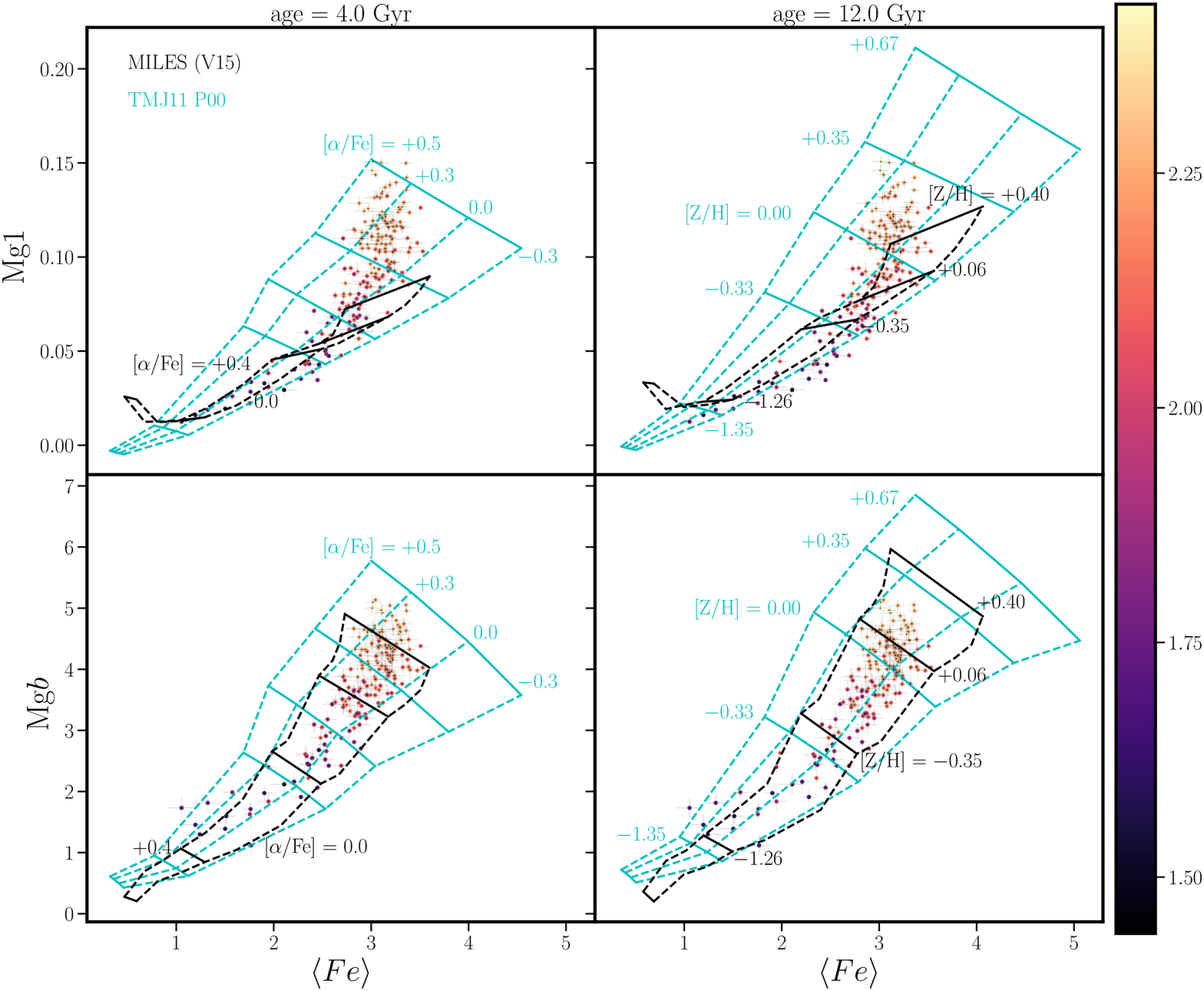}
    \caption{
    The model grids of V15 (black) and TMJ11 (cyan) in \mgI-$<$Fe$>$ and \mgb-$<$Fe$>$ planes (the upper and lower panels) at 4 and 12~Gyr (left and right columns). Beside the isolines we label the corresponding \ZH\ or \afe\ values. The data points of our sample ETGs are colour coded according to the log(\sgmReKms). The colour bar is displayed on the right. 
    At each \ZH, while the \mgI\ index has a higher value at a higher \afe\ in TMJ11 model, it behaves inversely in V15 models. By contrast, the \mgb\ index has higher values at higher \afe\ in both models.
     }
    \label{fig:Mg1MgbFe_grid}
\end{figure*}

In Section~\ref{sec:compMethods} we mention that \mgI\ has higher values in the scaled-solar models than in the \alf-enhanced models. It is contrary to the behaviours of \mgI\ in other stellar population models (e.g., TMJ11). 
In Figure~\ref{fig:Mg1MgbFe_grid} we compare the model grids of V15 and TMJ11. We plot the grids of these two models in black and cyan respectively. The upper panels display the grids in \mgI-$<$Fe$>$ planes. The \ZH\ and \afe\ values are labeled beside the corresponding isolines. We colour code the data points according to the log(\sgmReKms) of our sample ETGs. At each \ZH, while the \mgI\ index has a higher value at a higher \afe\ in TMJ11 model, it behaves inversely in V15 models. In the lower panels, we replace the \mgI\ with \mgb\ for y-axis. In this case, in both models, the \mgb\ index has higher values at higher \afe. 
The different behaviour of \mgI\ may be due to the different treatment of C in two models. From the Figure~1 in \citet{Johansson_12}, while \mgb\ is mostly sensitive to Mg abundance, \mgI\ is much more sensitive to the abundance of C comparing to Mg. 

It is possible that \mgI\ plays a role in shaping the \afe-\sgm\ relation. Therefore we compare the results from the fits with and without \mgI\ to examine the role of \mgI. 
In Section~\ref{sec:compSpecI} we apply the spectral fits with the wavelength coverage of only the 8 Lick indices within 4800-5500\AA. Here we exclude the \mgI\ index and repeat the fits on the same data we apply in Section~\ref{sec:compSpecI}. 

\begin{figure*}
	\includegraphics[width=\textwidth]{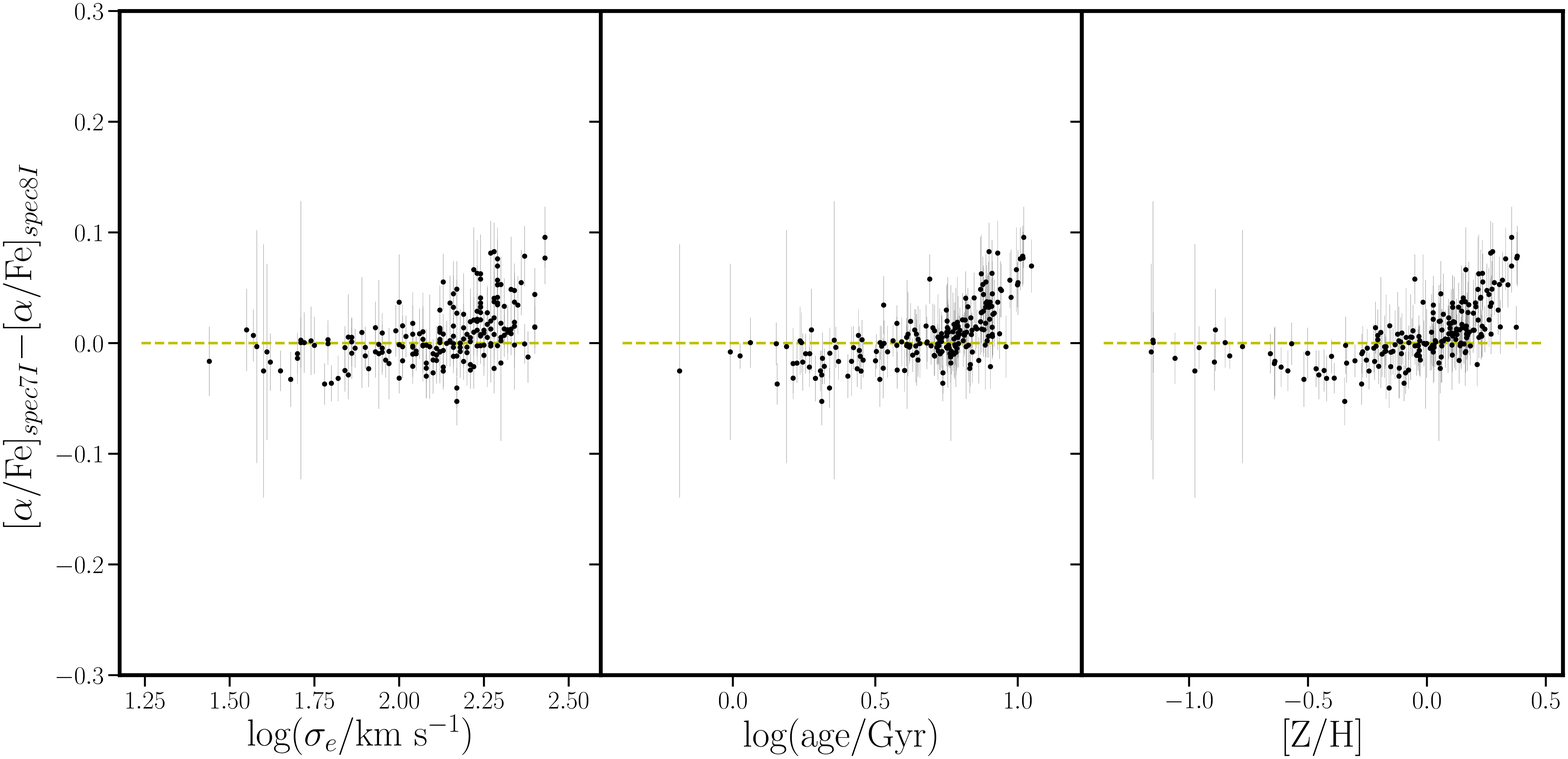}
    \caption{
    Similar to Figure~\ref{fig:dltV_afeSpec8I}. We apply the spectral fits with the wavelength coverage of the Lick indices within 4800-5500\AA, including or excluding the \mgI\ band. The y-axis indicates the difference of the \afe\ measurements from the fits with 7 and 8 Lick index bands, i.e., excluding and including the \mgI\ band. The difference is prominent at large \sgmRe, old ages, and high \ZH, and correlates with the log(age/Gyr) and \ZH\ best. 
     }
    \label{fig:dlt8I_afeSpecMg1}
\end{figure*}
\begin{figure}
	\includegraphics[width=\columnwidth]{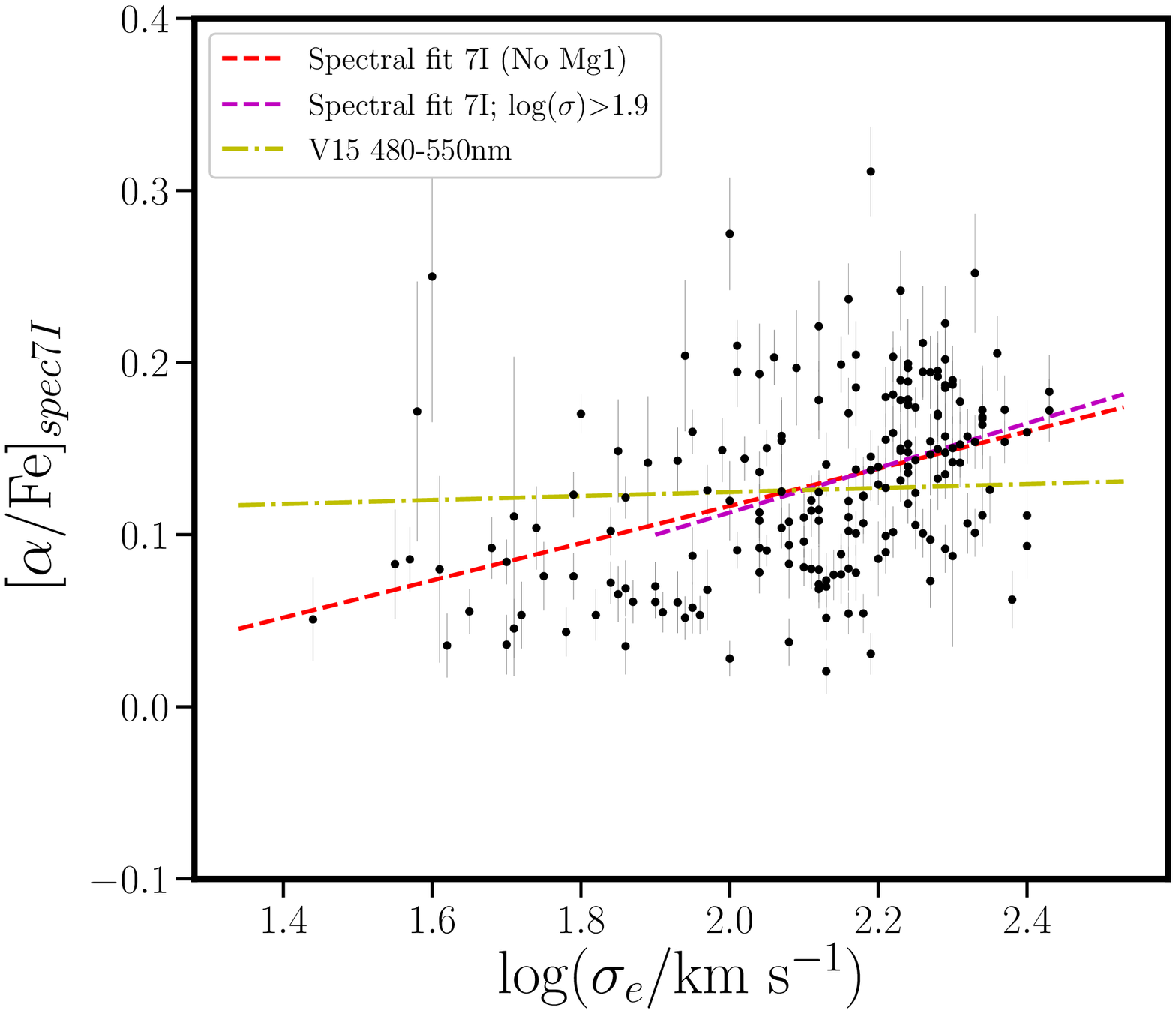}
    \caption{
    Similar to Figure~\ref{fig:afe_sgm}, but the y-axis is the \afe\ that is fitted with the wavelength coverage of 7 Lick indices within 4800-5500\AA, excluding the \mgI\ band. The red and magenta dashed lines are the relations fitted from the entire and massive samples. The yellow dot-dashed line is the relation from our main results (the magenta dashed line in Figure~\ref{fig:afe_sgm}). 
     }
    \label{fig:aFeSpec7I_sgm}
\end{figure}

We compare the \afe\ results from the fits over the bands of 7 and 8 Lick indices. Figure~\ref{fig:dlt8I_afeSpecMg1} shows the differences. While including \mgI\ or not does not change the \afe\ measurements at log(\sgmRe/km s$^{-1}$)$<$2.1, the difference increases with \sgm\ at the massive end. At super solar \ZH\, log(age/Gyr)>0.7, or log(\sgmReKms)$>$2.1, the differences have slightly tighter correlation with \ZH\ or log(age) than with log(\sgmRe). 
It implies that the \mgI-driven difference is related to \ZH. 
According to V15, the \alf-enhanced models at high metallicity incorporate theoretical differential spectral response to switch from the abundance pattern of the Milky Way to the desired enhancement. However, the computation of very cool star spectra, which contribute to the molecular bands like MgH is difficult, making these features less reliable for the analysis performed here. 
We suggest avoiding this rather wide index, which covers 471\AA, as it might suffer from other effects such as flux-calibration issues. 

Such offset of the \afe\ measurements at high-\ZH\ and massive ends significantly steepens the \afe-\sgm\ relation. In Figure~\ref{fig:aFeSpec7I_sgm} we plot the \afe\ measured from the spectral fits over the 7 indices against log(\sgmReKms). The red and magenta dashed lines are the relations fitted from the entire and massive samples, with the slopes of 0.1 and 0.13 respectively. The yellow dot-dashed line is the relation from our main results (the magenta dashed line in Figure~\ref{fig:afe_sgm}), plotted for reference. 

\begin{figure}
	\includegraphics[width=\columnwidth]{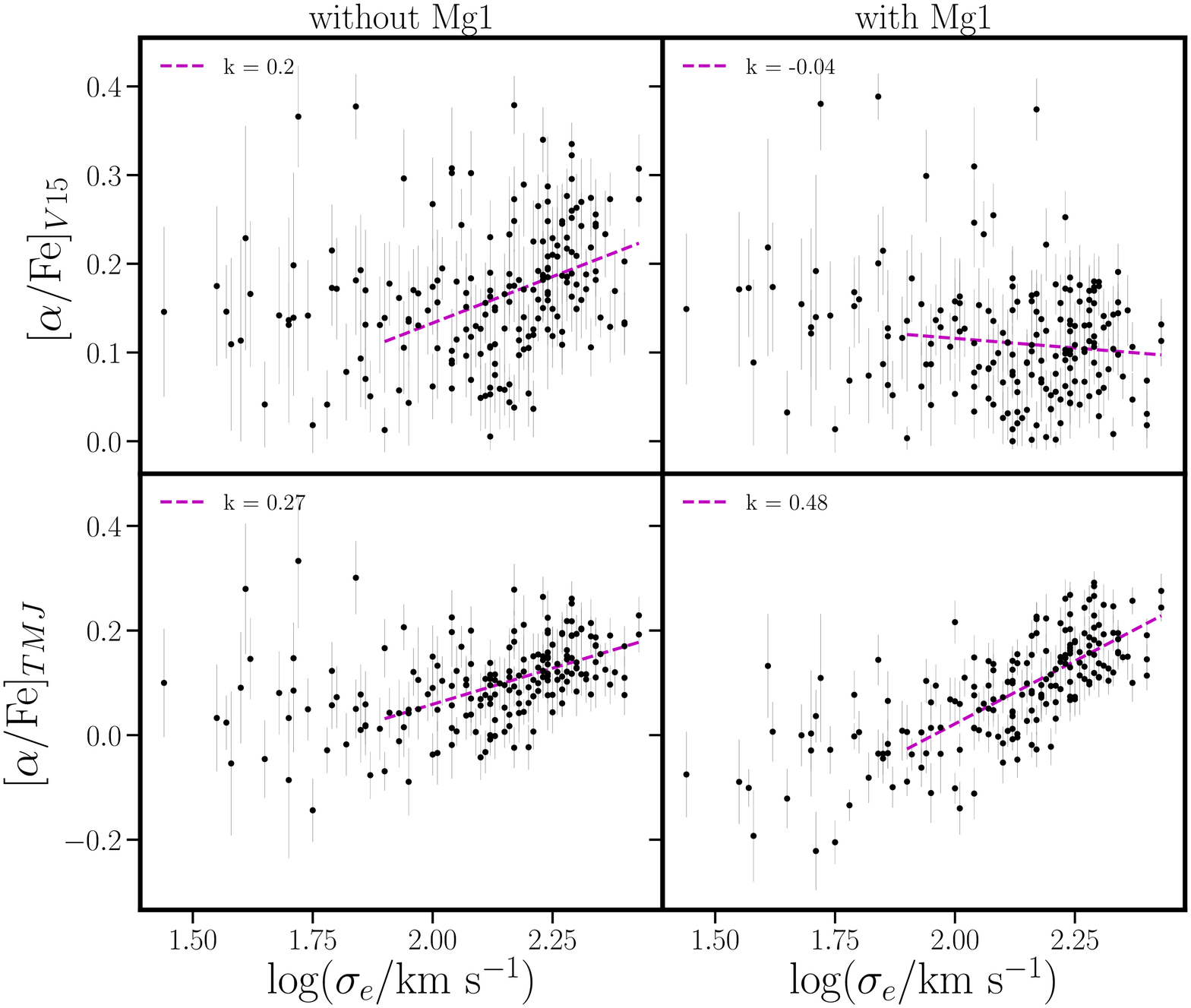}
    \caption{
    In the left column we plot the \afe\ measurements which are presented in Figure~\ref{fig:TMJ_V3I} against log(\sgmReKms). These \afe\ are derived from the fits with three indices, \mgb, [MgFe]', and Fe3, at fixed ages. The upper and lower panels correspond to the fits using V15 and TMJ11 models. The right column presents the \afe\ which are fitted by the same methods but with an additional index of \mgI. 
    The magenta dashed lines illustrate the linear relations that are fitted from the massive sample. The slopes are labeled in the upper left corners. 
    Without \mgI, the \afe\ fitted from both models has positive correlation with log(\sgmRe) in massive ranges, but the scatter is larger when using V15 models. 
    After including \mgI, while the \afe-\sgm\ relation of massive ETGs keeps positive when using TMJ11 model, the V15 models make it flat. 
     }
    \label{fig:compMg1_afe}
\end{figure}

In Figure~\ref{fig:compMg1_afe}, we show more direct comparisons. In Section~\ref{sec:Test:LickI}, we mention the Lick index approach which V15 suggested. They estimated the ``true \afe" by fitting \afe\ and \ZH\ with three indices \mgb, 
[MgFe]'=$\sqrt{Mgb\times(0.72Fe5270+0.28Fe5335)}$, and Fe3=(Fe4383+Fe5270+Fe5335)/3 at the fixed ages which were derived from the the spectral fits with V10 templates. We have repeated such analysis with both V15 and TMJ11 models on our sample. In Figure~\ref{fig:TMJ_V3I} we compare the \afe\ results measured from two models and the correlation is in agreement with V15's. 
In Figure~\ref{fig:compMg1_afe}, we plot these \afe\ results against log(\sgmReKms) in the left column. The upper and lower panels indicate the results from V15 and TMJ11 models respectively. The magenta dashed lines illustrate the linear relations that are fitted from the massive sample. In each panel, we label the slope of the relation in the upper left corner. 
The \afe\ fitted from both models has positive correlation with log(\sgmRe) in the massive ranges, but the scatter is larger when using V15 models. 

At the same time, we fit the \afe\ using the same methods but including \mgI\ in our fits in addition to those three indices. We show the results correspondingly in the right column. 
The inclusion of \mgI\ flattens the classic positive relation when using V15 models.
By contrast, using TMJ11 model, the positive \afe-\sgm\ relation of massive ETGs still holds after the inclusion of \mgI. The slope becomes even steeper. This is because from Figure~\ref{fig:Mg1MgbFe_grid}, our massive sample locate in the positions with higher \afe\ in \mgI-$<$Fe$>$ planes than in \mgb-$<$Fe$>$ planes. 
The different relations between log(\sgmRe) and the \afe\ fitted by different models with or without the inclusion of \mgI\ indicate that V15 models have difficulties in covering the observations in \mgI. 

\subsection{The \afe-\sgm\ Relation of ETGs}
\label{sec:aFe_sgm}

As mentioned in the introduction, all the previous stellar population analysis with other stellar population models show positive correlations between the global \afe\ and \sgm\ among massive ETGs. 
Although some simulations can produce the positive slopes \citep[e.g.,][]{Arrigoni_10,Yates_13,Segers_16,Vincenzo_18}, they all include special mechanisms, e.g., more top-heavy IMFs, different supernova properties, and AGN feedback. None of these mechanisms has been supported by robust evidence from observations. 

A potential reason for the discrepancy is that simulations and observations use different \alf\ element indicators. In simulations the most common \afe\ indicator is [{\rm O}/{\rm Fe}]. Comparing to [{\rm Mg}/{\rm Fe}] and [{\rm Si}/{\rm Fe}], the [{\rm O}/{\rm Fe}] distribution of the Milky Way stars matches the theoretical predictions better \citep{MatteucciGreggio_86}. However, comparing to Mg, it is more difficult to estimate the stellar O abundance using relatively low resolution spectra of the integral extragalactic light. Therefore in most stellar population analysis, the \alf\ elements are traced by Mg. 

In most previous studies, the adopted indicator of \alf\ elements is Mg$b$ index. From our analysis, the inclusion of \mgI\ makes difference in the \afe\ measurements of ETGs with super-solar \ZH. 
On one hand, according to Section~\ref{sec:mg1}, we suggest that it is a consequence of the self-inconsistency of V15 models. 
On the other hand, it indicates that we cannot rule out the probability that the \afe\ of massive ETGs is not correlated with their \sgm. In other words, though the probability is low, the galaxies with deeper potential wells might not have more intense early star formation processes, different IMFs, or special supernova properties.


\section{Conclusion}
\label{sec:Conclusion}

We measure the mean \afe\ and the \afe\ radial gradients within the 1\Rmaj\ ellipses of 196 high-S/N MaNGA ETGs. The \sgmRe\ ranges from 27 to 270~km/s. We use the \ppxf\ based spectral fitting methods. The templates are produced by the V15 models which are constructed on BaSTI scaled-solar and \alf-enhanced isochrones. We only adopt the wavelength range of 4800-5500~\AA\ in our fits. 

Our results of the galactic mean \afe\ are marginally correlated with galactic velocity dispersions, for both the entire sample and the massive ETGs. It is different from the classic positive \afe-\sgm\ relations of ETGs, especially in the high-mass range. The difference is not driven by our sample selection, because we successfully reproduce the positive \afe-\sgm\ relation when we estimate \afe\ by the traditional Lick index analysis with other SSP models. 

Including the non-prominent features beyond the Lick indices or not does not make much difference on \afe\ estimations. 
Comparing to the CSP spectral fits, fitting with SSP would result in higher \afe\ at low masses and relatively low \ZH. At \ZH$<-$0.3, the differences are larger at lower \ZH, probably due to the degeneracy between low \ZH\ and high \afe. At log(\sgmRe/km s$^{-1}$)$>$1.9 or \ZH$>-$0.3, however, the differences are small and do not show correlations with \sgmRe\ or \ZH. Therefore replacing the CSP fits with SSP fits does not make our \afe-\sgm\ relations positive. 

The primary reason for the flatness of our \afe-\sgm\ relations is the inclusion of \mgI\ band in our spectral fits using V15 model templates. 
We could recover the traditional positive \afe-\sgm\ relation of massive ETGs when excluding the \mgI\ feature in our fits. In V15 models, \mgI\ has higher values in the scaled-solar models than in the \alf-enhanced models. Therefore including the \mgI\ band results in lower \afe\ estimations for massive ETGs. We find that the difference of \afe\ measurements caused by \mgI\ correlates with \ZH\ and \sgmRe, and the correlation with \ZH\ is tighter. The difference is larger at higher \ZH\ or \sgmRe. 
In other stellar population models where \mgI\ does not have such behaviours, including \mgI\ or not will not flatten the \afe-\sgm\ relations. 

Our results of the \afe\ gradients are generally flat. 
At massive end, they have slightly positive gradients. These results indicate that the star formation timescale of ETGs does not vary much within their inner 1\re. For massive ETGs, they have slightly more extended star formation histories in the central regions. 
However, because of the \mgI\ issue, the flatness of our \afe\ radial profiles might be due to the possible underestimations of some intrinsically high \afe. 

For reference, we also measure the gradients of log(age) and \ZH. They generally have similar trends with log(\sgmRe), but the log(age) radial profiles are overall much flatter. Our log(age) and \ZH\ gradients decrease as \sgmRe\ increases. On average, they are positive at low masses then decrease to negative in the high-mass range. 
Due to the low resolution of the age estimations for old objects, we note that the uncertainties of the age gradient measurements cannot be neglected.

\section*{Acknowledgements}

Y.L. thanks the anonymous referee for the helpful advices on this paper. 
She appreciates the astronomers in the Astrophysics Sub-department at Oxford University, primarily professors Roger Davies and Michele Cappellari, as well as other colleagues working on galactic astronomy, e.g., Sam Vaughan, Mark Graham, Martin Bureau, Chiaki Kobayashi, Clotilde Laigle, Aprajita Verma, and especially, Alexandre Vazdekis, for the invaluable discussions on this work. 
She acknowledges the support from the Oxford Centre for Astrophysical Surveys, which is funded through generous support from the Hintze Family Charitable Foundation. 


\section*{Data Availability} 

Data available on request. The data underlying this article will be shared on reasonable request to the corresponding author. 




\bibliographystyle{mnras}
\bibliography{ref} 





%


\bsp	
\label{lastpage}
\end{document}